\documentclass[sn-basic]{sn-jnl}

\usepackage{graphicx}%
\usepackage{multirow}%
\usepackage{amsmath,amssymb,amsfonts}%
\usepackage{amsthm}%
\usepackage{mathrsfs}%
\usepackage[title]{appendix}%
\usepackage{xcolor}%
\usepackage{textcomp}%
\usepackage{manyfoot}%
\usepackage{booktabs}%
\usepackage{algorithm}%
\usepackage{algorithmicx}%
\usepackage{algpseudocode}%
\usepackage{listings}%
\usepackage{xcolor} 

\definecolor{keywordcolor}{rgb}{0.13,0.13,1}    
\definecolor{stringcolor}{rgb}{0.2,0.6,0.2}     
\definecolor{commentcolor}{rgb}{0.5,0.5,0.5}    

\lstdefinestyle{mystyle}{
  language=Python,
  basicstyle=\footnotesize\ttfamily,
  keywordstyle=\color{keywordcolor}\bfseries,
  stringstyle=\color{stringcolor},
  commentstyle=\color{commentcolor}\itshape,
  breaklines=true,
  columns=fullflexible,
  keepspaces=true,
  frame=single,
  rulecolor=\color{black},
  showstringspaces=false
}

\usepackage{tabularx}
\usepackage{nameref,hyperref,graphicx}
\usepackage{amsmath,amssymb}
\usepackage{rotating}
\usepackage{changepage}
\usepackage{caption}
\usepackage{subcaption}
\usepackage[table]{xcolor}
\usepackage{array}
\usepackage{geometry}

\theoremstyle{thmstyleone}%
\theoremstyle{thmstyletwo}%

\theoremstyle{thmstylethree}%

\newcommand{\methodname}{pyBiblioNet}

\begin{document}

\title[\methodname{}: A python library for a comprehensive network-based bibliometric analysis]{\methodname{}: A python library for a comprehensive network-based bibliometric analysis}


\author*[1]{\fnm{Mirko} \sur{Lai}}\email{mirko.lai@uniupo.it}
\equalcont{These authors contributed equally to this work.}

\author[1]{\fnm{Salvatore} \sur{Vilella}}\email{salvatore.vilella@uniupo.it}
\equalcont{These authors contributed equally to this work.}

\author[2]{\fnm{Federica} \sur{Cena}}\email{federica.cena@unito.it}

\author[1]{\fnm{Giancarlo} \sur{Ruffo}}\email{giancarlo.ruffo@uniupo.it}

\affil*[1]{\orgdiv{Dipartimento di Scienze e Innovazione Tecnologica}, \orgname{Università del Piemonte Orientale}, \orgaddress{\street{V.le Teresa Michel, 11}, \city{Alessandria}, \postcode{15121}, \state{Piedmont}, \country{Italy}}}

\affil[2]{\orgdiv{Dipartimento di Informatica}, \orgname{Università degli Studi di Torino}, \orgaddress{\street{C.so Svizzera 185, Via Pessinetto, 12}, \city{Turin}, \postcode{10149}, \state{Piedmont}, \country{Italy}}}


\abstract{Bibliometric analysis is a critical tool for understanding the structure, dynamics, and impact of scientific research. Traditional methods often fall short in capturing the intricate relationships and evolving trends within scientific literature. To address this gap, we present \methodname{}, a Python library designed to facilitate comprehensive network-based bibliometric analysis, providing insights into citation networks, co-authorship networks, and keyword co-occurrence networks. The library integrates with OpenAlex, a popular and open catalogue to the global research system, enabling users to easily preprocess, visualize, and analyse bibliometric data. Key features include topic selection, automatic data download via OpenAlex APIs, creation of the root and base sets of manuscripts to analyze, creation of the citation and co-authorship networks, network visualization tools, and a suite of algorithms for computing network centralities, clustering, and community detection, all of them tailored to the bibliometric domain. Additionally, it enables the analysis of key topics and concepts using NLP techniques. We showcase the main functions of the library by performing a bibliometric analysis on the multidisciplinary ``15-minute city paradigm'', demonstrating the utility of \methodname{} in uncovering hidden patterns and emerging trends in various scientific domains. \methodname{} can empower researchers, librarians, and policymakers with a powerful, user-friendly tool for enhancing their bibliometric analyses and making data-driven decisions.}

\keywords{bibliometric analysis, network analysis, python library, 15 minute city}



\maketitle

\section{Introduction and related work}

Bibliometric analysis is an essential methodology for exploring the structure, dynamics, and impact of scientific research. By systematically quantifying and analysing scholarly outputs, bibliometric techniques enable researchers to uncover patterns in citation practices, co-authorship relationships, and the evolution of research themes. While all scientific domains have their traditions, habits and common practices when it comes to bibliographic research, there exists a number of cross-domain tools \cite{moral2020software}, that help researchers in retrieving information, organize and explore their preferred bibliographic sources. The presence of such tools is becoming crucial, considering the ever-increasing growing rate of papers published every year, that saw an exponential surge since the 1990s~\cite{fire2019over}, making it considerably harder for scientists to keep up-to-date in their field of research, and even harder for them to approach a domain outside their expertise. 

Within this context, it might be instrumental to integrate new approaches to facilitate bibliometric tasks, such as including network analysis, which offers complementary metrics relevant to bibliometric research. As highlighted in~\cite{donthu2021conduct}, network analysis is an integral part of modern bibliometric studies: many literature surveys in a wide variety of fields have used it to uncover relationships between research domains and sub-domains, authors, research institutions and, most importantly, ideas and concepts, e.g.,~\cite{liu2015visualizing,andrikopoulos2016coauthorship,tunger2018bibliometric,rossetto2018structure,cisneros2018bibliometric,baker2020bibliometric,andersen2021mapping,ruffo2023studying}.

Indeed, by construction, network analysis allows for the examination of relationships among publications through various network types. Citation networks illustrate how publications are interconnected via citations, while collaboration networks reveal co-authorship patterns among researchers. Additionally, semantic networks can highlight common terms across publications, providing insights into thematic connections within the literature. Techniques such as co-citation and bibliographic coupling further enrich this analysis by identifying clusters of related works and emerging research areas. Tools like VOSviewer~\cite{van2010software}, which employs a distance-based approach to create visual representations of citation relationships and co-authorship networks, and Gephi~\cite{bastian2009gephi} are frequently employed to visualize these networks, offering graphical representations that enhance the understanding of complex bibliometric relationships~\cite{van2014visualizing}. This multifaceted approach not only aids in identifying influential authors and journals but also helps in mapping the evolution of research fields over time. Other solutions, such as Research Rabbit~\cite{researchrabbit_website}, provide bibliography management tools together with the visualization of citation and co-authorship networks.

The development of bibliometric tools has gained momentum in recent years, with several software libraries to facilitate data collection and analysis \cite{moral2020software}. Notable among these is Bibliometrix~\cite{aria2017bibliometrix}, an R package that provides extensive functionalities for bibliometric analysis, including data extraction from various databases and visualization capabilities. Another significant contribution is pyBibX~\cite{pereira2023pybibx}, which integrates artificial intelligence capabilities (including Embedding vectors, Topic Modeling, Text Summarization, and other general Natural Language Processing tasks, employing models such as Sentence-BERT, BerTopic, BERT, chatGPT, and PEGASUS) into bibliometric analysis by allowing exploratory data analysis and network capabilities such as citation and collaboration analysis.

In a broader context, several authors (e.g.,~\cite{Salman2025SystematicAO,Tosi2025ComparingGA}) have highlighted the growing need to accelerate research workflows through the adoption of automated tools that can reduce time spent on literature synthesis via reliable search and summarization methods. While generative AI tools like ChatGPT and Perplexity offer promising opportunities, they also raise ethical concerns, such as the risks of plagiarism, bias, and over-reliance on AI-generated content. Therefore, it is essential to combine the efficiency of AI with the critical thinking and oversight of human researchers. From this perspective, challenges related to interpretation and explainability become crucial and should not be neglected. Network-based bibliometric analysis offers a robust theoretical framework in which each metric can be clearly understood, explored, and visualized. We argue that, in the short term, network-based bibliometric analysis will complement rather than be replaced by generative AI tools, particularly in addressing the issue of information overload in scientific literature.

\subsection{Our contribution}

Such a diverse landscape of software solutions provides different answers to the need of facilitating bibliometric research \cite{moral2020software}. The most of them, like bibliometricx \cite{aria2017bibliometrix} and pybibx \cite{pereira2023pybibx}, include network analysis. However, it is often limited to visualization purposes, to obtain a general idea of the citation landscape and to identify cohesive clusters of papers and researchers; quite often they require manual data collection from source databases like Web of Science or Scopus, with data typically being downloaded in CSV or BibTeX format for further processing.

 We developed \methodname{}, a Python library specifically designed to enhance network-based bibliometric analysis. \methodname{} consists of two main modules: one devoted to data collection and one aimed at community detection and network analysis. 
 The first module  leverages OpenAlex~\cite{priem2022openalex} APIs to easily collect bibliometric data and construct a citation graph (see documentation of the module \texttt{openalex} in Appendix~\ref{appendixA}). While other data sources can be included, the graph object is the core of the analysis implemented in the second module, that performs network metrics calculations and identifies relevant nodes and connections. Furthermore, it focuses on community detection, implementing different popular algorithms, such as Louvain \cite{blondel2008fast}, Infomap \cite{rosvall2008maps}, Stochastic Block Models \cite{holland1983stochastic} (as detailed in Table \ref{tab:community_detection}), allowing the user to visualize the connections between clusters in a customizable visualization (see documentation of the module \texttt{bibliometric\_analysis} in Appendix~\ref{appendixB}). Since \methodname{} is released as open source software\footnote{\url{https://github.com/mirkolai/pybiblionet} and \url{https://pypi.org/project/pybiblionet/}}, it allows the users to integrate their own data sources, analyses and visualizations, making it an extremely flexible tool for network-based bibliometric research.

At the time of the release of the pyBibX library, Pereira et al. \cite{pereira2023pybibx} conducted an in-depth analysis of 15 different bibliometric tools, providing for each of them a detailed summary of key features, data sources, and operational environments.
In this article, we compare our system, \methodname{},  with the three tools that, according to their analysis, satisfied the highest number of identified key features, i.e. pyBibX \cite{pereira2023pybibx}, Bibliometrix \cite{aria2017bibliometrix}, and Scientopy \cite{Scientopy2019}.
Table \ref{tab:comparison} presents a comparative analysis between \methodname{}, pyBibX, Bibliometrix, and Scientopy.

\begin{table}[ht]
\footnotesize
\centering
\begin{tabular}{@{}p{1.6cm}p{1.5cm}p{2.5cm}p{3cm}p{2.8cm}@{}}
\toprule
Tool         & Source                                    & Data Analysis                                                                                                                         & Network Analysis                                                                                                                    & Artificial Intelligence                                      \\ \midrule
bibliometrix & \begin{tabular}[c]{@{}l@{}}Cochrane\\ Dimensions\\ PubMed\\ Scopus\\ WoS\end{tabular} & \begin{tabular}[c]{@{}l@{}}Data Manipulation\\ Data Analysis\\ Wordcloud\\ Projection\\ Evolution Plot\\ Sankey Diagram\\ Bar Plot\end{tabular} & \begin{tabular}[c]{@{}l@{}}Citation Analysis\\ Collaboration Analysis\\ Similarity Analysis\\ World Collab. Analysis\end{tabular} & -                                                          \\\hline
pyBibX       & \begin{tabular}[c]{@{}l@{}}PubMed\\ Scopus\\ WoS\end{tabular}                       & \begin{tabular}[c]{@{}l@{}}Data Manipulation\\  Data Analysis\\ Wordcloud\\ N-Gram\\ Projection\\ Evolution Plot\\ Sankey Diagram\\ Treemap\\ Bar Plot\end{tabular} & \begin{tabular}[c]{@{}l@{}}Citation Analysis\\ Collaboration Analysis\\ Similarity Analysis\\ World Collab. Analysis\end{tabular} & \begin{tabular}[c]{@{}l@{}}Topic Modeling\\ Embeddings\\ Summarization\\ chatGPT \end{tabular} \\\hline
Scientopy    & \begin{tabular}[c]{@{}l@{}}Scopus\\ WoS\end{tabular}                               & \begin{tabular}[c]{@{}l@{}}Data Manipulation\\  Data Analysis\\ Wordcloud\\ Evolution Plot\\ Bar Plot\end{tabular}                         & \begin{tabular}[c]{@{}l@{}}Citation Analysis\\ Collaboration Analysis\\ Similarity Analysis\\ World Collab. Analysis\end{tabular}               & -    
  \\\hline  
  pyBiblioNet  & OpenAlex                                  & \begin{tabular}[c]{@{}l@{}}Clustered Network\\ N-Gram\\ Evolution Plot\\ Bar Plot\end{tabular}                                                     & \begin{tabular}[c]{@{}l@{}}Citation Analysis\\ Collaboration Analysis\\ Cluster Analysis\\ Base Set Analysis\\ Centrality Analysis \end{tabular}     & Topic Modeling                                                                              \\ \bottomrule
\end{tabular}
\caption{Comparison of bibliometric tools in terms of source, data analysis capabilities, network analysis features, and AI integration.}
\label{tab:comparison}
\end{table}

Among the analysed libraries, pyBibX offers the highest number of features, including advanced features that employ large language models.
With pyBiblioNet, our objective was not to replicate functionalities already implemented by existing libraries, but rather to provide novel and distinctive features.

First of all, pyBiblioNet is the only tool that utilizes OpenAlex as a data source that is not only an alternative but also integrates multiple data sources, including ORCID, ROR, DOAJ, Unpaywall, PubMed, PubMed Central, and the ISSN International Center. Additionally, it offers a significantly larger number of articles (243 million) compared to Scopus (87 million), Web of Science (Core) (87 million), and Dimensions (135 million)\footnote{\url{https://openalex.org/about\#comparison}}.

Furthermore, although our current analysis focuses primarily on citation and co-authorship networks, we introduced the possibility of expanding the network by including base set nodes — that is, articles that do not initially satisfy the search filters but are connected to the root set network — thereby broadening the scope of the analysis.
Another important innovation is the application of clustering techniques to the citation network in order to perform thematic analysis of groups of articles.
A similar clustering approach can be applied to the co-authorship network to study collaboration patterns between countries.
Finally, we made centrality metrics easily accessible, allowing for the identification of influential articles based on criteria beyond simple citation counts, such as the number of collaborations.

In order to avoid creating yet another isolated library, we also aimed to ensure interoperability with existing tools by allowing the export of metadata retrieved from OpenAlex in a format compatible with other libraries. Moreover, citation and co-authorship networks can be exported in standardized formats suitable for use with external network analysis tools.

\section{Network-based bibliometrics}

As pointed out in the previous section, network-based bibliometrics is a set of analytical methods and tools that employs network science to study and quantify the relationships and influence among scientific entities, such as authors, articles, journals, or institutions. This methodology is rooted in the principles of graph theory and social network analysis, where scientific bibliometric networks are often represented as \textit{directed or undirected graphs} and \textit{weighted or unweighted graphs}, with scientific entities represented as \textit{nodes} connected by \textit{edges} that denote interactions between them. A \textit{directed graph} \( G = (V, E) \) consists of a set of nodes \( V \) and a set of directed edges \( E \), where each edge has a direction, denoted by an ordered pair \( (u, v) \in E \), meaning that there is a directed connection from node \( u \) to node \( v \). 
Conversely, an \textit{undirected graph} is one in which the edges have no direction, and the connection between two nodes \( u \) and \( v \) is represented simply by an unordered pair \( \{u, v\} \).
This representation is appropriate when the relationship between entities is mutual, such as co-authorship between researchers. 
Additionally, a \textit{weighted graph} is a graph in which each edge is assigned a weight, typically representing the strength or intensity of the relationship between two nodes. 
Mathematically, a weighted graph can be represented as \( G = (V, E, w) \), where \( w: E \to \mathbb{R} \) is a weight function that assigns a real-valued weight to each edge. 
For bibliometric networks, these weights might reflect, for instance, the number of collaborative works between two authors. 
An \textit{unweighted graph}, on the other hand, assumes that all relationships are equal and does not differentiate between the strengths of connections; for instance, this is the case for citation networks among articles, since an article can cite another only once.

\subsection{Centrality metrics}
The power of network analysis lies in the ability to identify \textit{key} actors in a simple and quantitative way. This can be achieved by computing a wide variety of centrality metrics, some of which are implemented in the library and listed in Tab.~\ref{tab:centralities}. These metrics go beyond the simple count of citations of a paper or the h-index of an author, quantifying in a more sophisticated way the importance of an entity in different roles. Some entities can be widely popular across different scientific communities or, on the contrary, can be extremely popular but only in their sub-domains; some can have a key role in bridging different communities, allowing researchers to transition from one group of papers to another; some have a tendency to be cited by - or to collaborate with - very authoritative entities. These and other nuances can be captured and quantified by network analysis, and are effective within the context of bibliometric analysis, as shown in Diallo et. al ~\cite{diallo2016identifying}.

\begin{table}[ht!]
\centering
\footnotesize
\begin{tabular}{|l|p{4.3cm}||p{4.5cm}|p{4.5cm}|}
\hline
\textbf{Metric} & \textbf{Definition} & \textbf{Relevance} \\ \hline
Degree Centrality & $C_D(v) = \deg(v)$, where $\deg(v)$ is the number of edges incident to node $v$. 

& Identifies frequently cited papers or prolific authors, indicating direct influence or productivity. \\ \hline
Betweenness Centrality & $C_B(v) = \sum_{s \neq v \neq t} \frac{\sigma_{st}(v)}{\sigma_{st}}$, where $\sigma_{st}$ is the total number of shortest paths from $s$ to $t$, and $\sigma_{st}(v)$ is the number of those paths passing through $v$. 

& Highlights nodes serving as intermediaries, facilitating the flow of information across different parts of the network, such as interdisciplinary papers. \\ \hline
Closeness Centrality & $C_C(v) = \frac{1}{\sum_{u \neq v} d(v,u)}$, where $d(v,u)$ is the shortest path distance between $v$ and $u$. 

& Identifies nodes that are strategically positioned to efficiently access or disseminate information throughout the network. \\ \hline
Eigenvector Centrality & $C_E(v) \propto \sum_{u \in V} A_{vu} C_E(u)$, where $A$ is the adjacency matrix of the network. 

& Identifies nodes with broader systemic impact, such as seminal papers cited by other highly influential works. \\ \hline
PageRank & $PR(A) = \frac{1 - d}{N} + d \sum_{i=1}^{k} \frac{PR(B_i)}{L(B_i)}$
where \( d \) is the damping factor, \( N \) is the total number of pages in the network, \( B_i \) are the pages linking to page \( A \), and \( L(B_i) \) denotes the number of outbound links from page \( B_i \).
. 

& Highlights authoritative or impactful works, even if they do not have the highest direct citation count. \\ \hline
\end{tabular}
\caption{Popular network-based centrality metrics and their relevance to bibliometric networks.}
\label{tab:centralities}
\end{table}

\subsection{Community detection }\label{subsec:comdet}

Community detection provides a powerful approach to transcending traditional categorizations based on scientific domain or discipline.  It reveals the natural structure of the system, uncovering the intrinsic organization of scientific knowledge and cooperation. Traditional classifications, such as those based on predefined scientific fields or subject areas, are often constrained by rigid boundaries and may fail to capture the dynamic and interdisciplinary nature of modern research. Community detection, on the other hand, identifies clusters of tightly connected nodes based on the actual relationships within the network; be they citation links, co-authorship ties, or other forms of interaction. These clusters represent organic groups of research activity, reflecting how knowledge is naturally organized and interconnected.

By detecting these communities, we gain insight into the true topology of the scientific landscape. Differently from  topic modelling, which extracts semantic themes from textual content, grouping documents by word usage, community detection relies on structural relationships (citations or co-authorships) to identify clusters of closely connected entities. Thus, it captures the structural and collaborative patterns of scientific activity, rather than purely thematic similarity. The two approaches are complementary, offering distinct but overlapping perspectives on how knowledge is produced and organized. For example, instead of viewing research on artificial intelligence as a monolithic field, community detection might reveal distinct subfields such as machine learning, natural language processing, and computer vision, as well as their connections to related areas such as neuroscience or ethics. Similarly, interdisciplinary research areas, which often fall between the cracks of traditional categorization, can emerge as distinct communities, highlighting their unique contribution to the scientific ecosystem. One can also study how communities evolve over time, identifying emerging fields or shifts in research focus. For instance, a growing cluster within a citation network might indicate an area of rapid innovation or increasing interest, providing early signals for trends in scientific research. Additionally, the ability to detect bridge nodes (i.e., those that connect distinct communities) enables the identification of key actors or works that facilitate knowledge transfer and interdisciplinary collaboration. 

There exists a large number of methodologies for community detection that can be applied to a wide variety of fields, with different perks and disadvantages. While we refer the reader to some scientific surveys for a comprehensive overview of the field~\cite{fortunato2010community, berahmand2023comprehensive, jin2024community}, we report a brief overview of some popular techniques, all of which are implemented in the library, in Tab.~\ref{tab:community_detection}, along with their relevance to bibliometric analysis.

\begin{table}[h!]
\centering
\begin{tabularx}{\linewidth}{|l|X|X|}
\hline
\textbf{Method} & \textbf{Definition} & \textbf{Relevance} \\ \hline
Louvain~\cite{blondel2008fast} & A modularity-based method that iteratively optimizes community structure for maximum modularity. & Efficient for analyzing large bibliometric networks, identifying clusters such as research disciplines or topical subfields based on modular structure. \\ \hline
Girvan-Newman~\cite{girvan2002community} & Uses edge betweenness to iteratively remove the most "central" edges, fragmenting the network. & Useful for identifying hierarchical structures in citation or co-authorship networks, such as nested subfields or interdisciplinary bridges. \\ \hline
Infomap~\cite{rosvall2008maps} & Partitions networks to minimize the description length of random walks. & Captures natural groupings in citation or collaboration networks, revealing how knowledge flows between topics or communities. \\ \hline
Spectral Clustering~\cite{ng2001spectral} & Partitions networks using the eigenvalues of graph Laplacians to identify clusters. & Suitable for detecting broad disciplinary boundaries or high-level structural patterns in bibliometric networks. \\ \hline
Stochastic Block Models~\cite{holland1983stochastic} & Assumes communities are latent variables generating the observed network structure. & Effective for probabilistic modeling of bibliometric networks. \\ \hline
\end{tabularx}
\caption{Popular methods for community detection and their relevance to bibliometric networks.}
\label{tab:community_detection}
\end{table}

\section{\methodname{}: Exploring bibliometric data through network analysis}

\begin{figure}[htb!]
    \centering
    \includegraphics[width=0.4\linewidth]{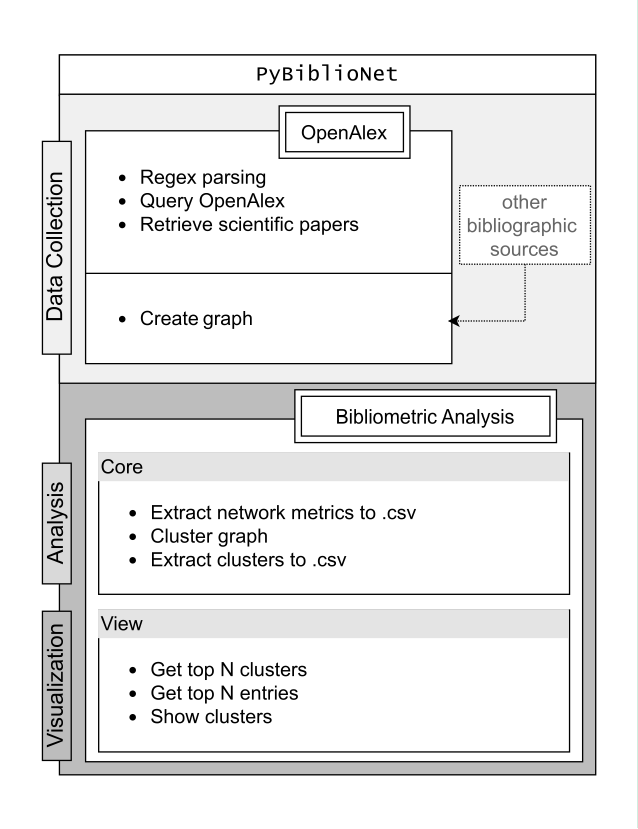}
    \caption{Scheme of \methodname{}. The library is divided into two main building blocks: a first block for data collection, that currently gathers data from OpenAlex but that can potentially be integrated by users with other bibliographic sources; the final output is a citation graph, that will be explored using the functions included in the second building block, devoted to bibliometric analysis.}
    \label{fig:scheme}
\end{figure}

The structure of \methodname{} is synthesized in Fig.~\ref{fig:scheme}. The structure consists of two main components: data collection and bibliometric analysis. The bibliometric analysis module is composed of the core analysis functionality and visualization modules, which enable the generation of graphs to facilitate interpretation. In this section, we will analyze the main components of the software.

\subsection{Root set/base set structure of citation networks}

The methodology underlying \methodname{} is based on the analysis of network metrics and on the concept of \textbf{root set/base set} in networks, as introduced in the HITS (Hyperlink-Induced Topic Search) algorithm developed by Jon Kleinberg in 1999 \cite{Kleinberg1999}. Broadly speaking, the root set represents the densely connected and most relevant part of the graph, typically including nodes central to the query. The base set instead includes less connected nodes, often on the network's fringes, such as articles cited by root set articles but not central themselves. Including the base set allows for a more comprehensive graph that provides insights into indirect relationships and broader connections, while excluding it focuses the analysis on the root set, often leading to a more compact and analytically focused network. The pseudocode \ref{alg:fetch_create_graph} illustrates the process.

It consists of two main steps: \texttt{FetchArticles} and \texttt{CreateCitationGraph}.
\texttt{FetchArticles} retrieves the \textbf{root set} and the base set articles from a bibliographic database (e.g., OpenAlex) based on user-defined search criteria. 
For each retrieved article, various metadata are available and the list of articles it cites and the list of articles that cite it are additionally included.
This creates a concise dataset representing a specific topic or area of interest.
The data collection process may require multiple queries, primarily because it needs to retrieve the list of cited or citing articles, and a considerable amount of time due to a predefined timeout set to prevent excessive load on the APIs. Additionally, OpenAlex allows up to 100,000 requests per user per day. Therefore, it is recommended to cache queries by setting the parameter \texttt{cache=True}.

\texttt{CreateCitationGraph} transforms the retrieved articles into a graph representation, where nodes correspond to articles and edges represent citation relationships. 
If \texttt{base set} is set to \texttt{False}, the graph includes only those articles matching the search criteria (the \textbf{root set} articles). Conversely, when \texttt{base set} is \texttt{True}, the graph is expanded to include citations involving articles outside the \textbf{root set}, creating a richer, context-aware graph that highlights the interactions between the root set and base set nodes.
In citation networks, the \textbf{root set} represents a tightly connected group of articles central to a specific topic or field. These articles are often highly cited and interrelated. The \textbf{base set}, on the other hand, consists of less-connected articles that may cite or be cited by the root set but are not central to the network. Including the base set provides a broader view of the citation landscape, which can reveal the influence of the root set articles on external works and vice versa.

Similarly, regarding the co-authorship network, the \texttt{CreateCoAuthorshipGraph} procedure allows the creation of an undirected network where nodes represent authors, and an edge exists when two authors co-publish an article. When \texttt{base set} is \texttt{True}, all articles are considered; otherwise, only the co-authorship relationships of the \texttt{root set} articles are included.

As an example, suppose a researcher is studying the field of "network science". The \texttt{FetchArticles} procedure retrieves articles strictly focused on this field, such as foundational papers and key recent studies. Setting \texttt{base set} to \texttt{False} in, the \texttt{CreateCitationGraph} procedure then builds a graph where nodes are these root set articles and edges represent citations between them, forming a dense, topic-specific network.

If the researcher instead sets \texttt{base set} to \texttt{True}, the graph is extended to include base set articles that cite the root set literature or are cited by it. For instance, articles from related fields like "social network analysis" or "complex systems" might appear in the graph. This broader perspective could reveal interdisciplinary influences and connections between network science and adjacent domains.

\begin{algorithm}[tp!] 
\scriptsize
\caption{Pseudocode that illustrates the process.}
\label{alg:fetch_create_graph}

\textbf{Fetch Articles:}
\begin{algorithmic}[1]
    \Procedure{FetchArticles}{$search\_criteria$}
        \State \texttt{articles} $\gets$ \texttt{fetch\_from\_openalex(search\_criteria)}
        \For{\texttt{article} in \texttt{articles}}
            \State \texttt{cited\_by} $\gets$ \texttt{get\_in\_citations(article)}
            \State \texttt{article.cited\_by} $\gets$ \texttt{cited\_by}
        
            \State \texttt{cite} $\gets$ \texttt{get\_out\_citations(article)}
            \State \texttt{article.cite} $\gets$ \texttt{cite}
        \EndFor
        \State \textbf{return} \texttt{articles}
    \EndProcedure
\end{algorithmic}
\vspace{0.2cm}
\textbf{Create CoAuthorship Graph:}
\begin{algorithmic}[1]
    \Procedure{CreateCoAuthorshipGraph}{$articles, baseset$}
        \State \texttt{graph} $\gets$ \texttt{initialize\_empty\_graph()}
        \State \texttt{done = set()}
        \For{\texttt{article in articles}}
            \State \texttt{done.add(article)}
            \For{\texttt{author1 in article}}
                    \State \texttt{add\_node\_to\_graph(graph, author1)}
                \For{\texttt{author2 in article}}
                    \State \texttt{add\_node\_to\_graph(graph, author2)}
                    \State \texttt{add\_edge\_to\_graph(graph, author1, author2)}
               \EndFor
            \EndFor
            \If{ \texttt{baseset} == \texttt{True}}

                \For{\texttt{article in article.cited\_by $\cup$ article.cite}}
                    \If{ \texttt{article} $\notin$= \texttt{done}}
                        \State \texttt{done.add(article)}
                        \For{\texttt{author1 in article}}
                            \State \texttt{add\_node\_to\_graph(graph, author1)}
                        \For{\texttt{author2 in article}}
                            \State \texttt{add\_node\_to\_graph(graph, author2)}
                            \State \texttt{add\_edge\_to\_graph(graph, author1, author2)}
                       \EndFor
                       
                    \EndFor
                        
                    \EndIf
        
                \EndFor
            \EndIf
        \EndFor

        \State \textbf{return} \texttt{graph}
    \EndProcedure
\end{algorithmic}
\vspace{0.2cm}

\textbf{Create Citation Graph:}
\begin{algorithmic}[1]
    \Procedure{CreateCitationGraph}{$articles, baseset$}
        \State \texttt{graph} $\gets$ \texttt{initialize\_empty\_graph()}
        \For{\texttt{article in articles}}
            \State \texttt{add\_node\_to\_graph(graph, article)}
        \EndFor

        \For{\texttt{article in articles}}
            \For{\texttt{cited\_by in article.cited\_by}}
                \If{ \texttt{baseset} == \texttt{False}}
                    \If{\texttt{cited\_by in articles}} 
                        \State \texttt{add\_edge\_to\_graph(graph, cited\_by, article)}
                    \EndIf
                \Else
                    \If{\texttt{cited\_by $\notin$ graph}} 
                        \State \texttt{add\_node\_to\_graph(graph, cited\_by)}
                    \EndIf
                    \State \texttt{add\_edge\_to\_graph(graph, cited\_by, article)}
                \EndIf
            \EndFor

            \For{\texttt{cite in article.cite}}
                \If{ \texttt{baseset} == \texttt{False}}
                    \If{\texttt{cite in articles}} 
                        \State \texttt{add\_edge\_to\_graph(graph, article, cite)}
                    \EndIf
                \Else
                    \If{\texttt{cite $\notin$ graph}} 
                        \State \texttt{add\_node\_to\_graph(graph, cite)}
                    \EndIf
                    \State \texttt{add\_edge\_to\_graph(graph, article, cite)}
                \EndIf
            \EndFor
        \EndFor

        \State \textbf{return} \texttt{graph}
    \EndProcedure
\end{algorithmic}

\end{algorithm}

\subsection{Data Collection: querying the OpenAlex API}
\methodname{} performs bibliometric analyses using data sourced from OpenAlex~\cite{priem2022openalex}. OpenAlex is an open-access database designed to provide comprehensive and structured information on scholarly works, including metadata on academic publications, authors, institutions, journals, and research concepts. It serves as a freely available alternative to proprietary academic databases such as Scopus and Web of Science. The data in OpenAlex is sourced from multiple repositories, including CrossRef, ORCID, and PubMed, ensuring broad coverage and regular updates. The database is organised in entities, which are summarised in Tab.~\ref{tab:openalex_entities}; these entities are interconnected, allowing for complex queries and relational analyses. OpenAlex offers a RESTful API for programmatic access to this data, facilitating its integration into bibliometric tools and analyses. By exploiting the API's functionalities, the user can perform complex boolean queries on multiple conditions, filtering results, searching for keywords and phrases and so on.

\begin{table}[h!]
\centering
\begin{tabular}{|l|p{10cm}|}
\hline
\textbf{Entity} & \textbf{Description} \\ \hline
Works & Individual scholarly outputs like journal articles, books, and conference papers. \\ \hline
Authors & Details on researchers, including affiliations and publication history. \\ \hline
Sources & Journals, conferences, and other publication outlets. \\ \hline
Institutions & Universities, research organizations, and other affiliated bodies. \\ \hline
Topics & Topics and research areas, allowing for topic modeling and analysis. \\ \hline
Publishers & List of publishers and related information. \\ \hline
Funders & List of research funders, with information coming from Crossref, and enhanced with data from Wikidata and ROR. \\ \hline
Geo & Information to geo-categorize scholarly data, using United Nations data to divide the globe into continents and regions to make data filtering easier. \\ \hline

\end{tabular}
\caption{Overview of OpenAlex Database Entities}
\label{tab:openalex_entities}
\end{table}

Just to mention a few examples, \texttt{https://api.openalex.org/works?search=dna} queries the \textit{works} entity searching for works with the search term "dna" in the title, abstract, or fulltext; \texttt{https://api.openalex.org/authors?filter=display\_name.search:einstein} searches in the \textit{authors} entity for all authors who have "Einstein" as part of their name; \texttt{https://api.openalex.org/works?filter=title.search:cubist} queries the \textit{works} entity searching for works with "cubist" specifically in the title; \texttt{https://api.openalex.org/works?search=(elmo AND "sesame street") NOT (cookie OR monster)} searches for \textit{works} that mention "elmo" and "sesame street," but not the words "cookie" or "monster".

Each one of these queries returns a JSON object with the results; each result is composed of many fields, containing a great deal of information on both the authors and the works. For a complete description of the available fields, we refer the reader to the OpenAlex API technical documentation\footnote{https://docs.openalex.org/how-to-use-the-api/api-overview, last accessed: 15/05/2025}.

As for the structure of the library, the module that handles data collection and storage includes functions to:

\begin{itemize}

    
    \item \textbf{interact with bibliographic APIs}. Through a custom wrapper, the user can interact with the OpenAlex APIs, streamlining the process of retrieving and analyzing scholarly data. Most importantly, it allows to specify the type of query as \texttt{'search'}, \texttt{'cite'}, or \texttt{'cited\_by'}, as well as to select publication-specific time windows. The function conveniently implements caching to reduce redundant API calls and avoid hitting rate limits. Such function, depending on the query type, will yield results in JSON format as specified in Tab.~\ref{tab:query_types}. 

    \begin{table}[h!]
    \centering
    \footnotesize
    \begin{tabular}{@{}lp{0.35\linewidth}p{0.45\linewidth}@{}}
    \toprule
    \textbf{Query Type} & \textbf{OpenAlex's API} & \textbf{Returns} \\ \midrule
    \texttt{search}     & \textit{works?filter=title\_and\_abstract.search:} &Searches the \textit{works} entity, looking for the specific query in titles and abstracts. \\
    \texttt{cite}       & \textit{works?filter=cites:} &Retrieves all the works that explicitly cite the queried parameter. \\
    \texttt{cited by}   & \textit{works?filter=cited\_by:} &Retrieves all the works that are cited by a specific article. \\ \bottomrule
    \end{tabular}
    \caption{\methodname{} query types, the exploited OpenAlex's API, and their respective returns.}
    \label{tab:query_types}
    \end{table}

    
    \item \textbf{Retrieving articles and constructing the graph}
    The function \texttt{retrieve\_articles} enables users to perform different types of queries, such as searching for articles, retrieving articles that cite a specific work, or finding works cited by a given article. The function handles the complexities of constructing API requests and parsing responses (Table \ref{tab:pybiblionet_functions_retriving}.
    To facilitate complex search queries, the wrapper includes a utility function, \texttt{string\_generator\_from\_lite\_regex}, which generates a list of query strings from a simplified regular expression pattern. This allows for efficient generation of multiple search queries from a pattern of alternatives.
    the \texttt{create\_citation\_graph} function creates a directed graph from the retrieved articles, where nodes represent articles and edges represent citation relationships.
    In addition, the \texttt{create\_coauthorship\_graph} function creates an undirected graph from the retrieved articles, where nodes represent authors ors, and an edge exists when two authors co-publish an article. These graphs can be used to visualize and analyze the structure of citation networks. These functions leverage the NetworkX library~\cite{hagberg2008exploring} to build the graph, which can also be exported in GML format for further analysis in graph analysis tools. The export has been specifically designed to ensure compatibility with the Gephi tool (version 0.10.1), allowing the use of partition options (e.g., language, article type, countries, etc.) in the appearance menu.
    Here, the users will also specify whether they are interested in including only the \textit{root set} or also the \textit{base set} articles into the graph.
        
    \begin{table}[h!]
    \centering
    \footnotesize
    \renewcommand{\arraystretch}{1.3} 
    \begin{tabularx}{\linewidth}{>{\raggedright\arraybackslash}p{0.25\linewidth}>{\raggedright\arraybackslash}p{0.2\linewidth}>{\raggedright\arraybackslash}X}
    \hline
    \textbf{Method} & \textbf{Task} & \textbf{Description} \\ 
    \hline
    \texttt{retrieve\_articles} & Interact with the bibliographic database & Fetches articles based on search queries, with options for retrieving citations and exporting the results to a CSV file. \\ \hline
    \texttt{create\_citation\_graph} & Constructing Citation Networks & Builds a directed graph of articles and their citation relationships, leveraging NetworkX for visualization and analysis. \\ \hline
    \texttt{create\_coauthorship\_graph} & Constructing Coauthorship Networks & Builds an undirected graph of authors and their co-authorship relationships, leveraging NetworkX for visualization and analysis. \\ \hline
    \end{tabularx}
    \caption{Overview of \methodname{} public methods employed for constructing the citation or a co-authorship Graph.}
    \label{tab:pybiblionet_functions_retriving}
    \end{table}

    \item \textbf{export the articles and authors' metadata}: 
    The functions \texttt{export\_articles\_to\_csv} and \texttt{export\_authors\_to\_csv} allow exporting the metadata of the retrieved articles and authors in a CSV file (Table \ref{tab:pybiblionet_functions_exporting}). Users can also specify which fields they wish to export and whether they want to include only the \textit{root set} articles or also the \textit{base set} articles.
    The library also supports the export of article venue information and author affiliations through the methods \texttt{export\_venue\_to\_csv} and \texttt{export\_institution\_to\_csv}.
    Finally, the function \texttt{export\_article\_to\_scopus} allows for metadata export, including article-level metadata in a Scopus-compatible CSV format, thereby enabling the seamless use of data retrieved through the OpenAlex wrapper as input for other bibliometric analysis tools.

    \begin{table}[h!]
    \centering
    \footnotesize
    \renewcommand{\arraystretch}{1.3} 
    \begin{tabularx}{\linewidth}{>{\raggedright\arraybackslash}p{0.25\linewidth}>{\raggedright\arraybackslash}p{0.2\linewidth}>{\raggedright\arraybackslash}X}
    \hline
    \textbf{Method} & \textbf{Task} & \textbf{Description} \\ 
    \hline
    \texttt{export\_article\_to\_csv} & Export metadata & Export articles metadata in CSV format. \\ \hline
    \texttt{export\_author\_to\_csv} & Export metadata  & Export authors metadata in CSV format. \\ \hline
    \texttt{export\_institution\_to\_csv} & Export metadata & Export institutions metadata in CSV format. \\ \hline
    \texttt{export\_venue\_to\_csv} & Export metadata  & Export venue metadata in CSV format. \\ \hline
    \texttt{export\_article\_to\_scopus} & Export metadata & Export articles' metadata in Scopus-like CSV format . \\ \hline
    \end{tabularx}
    \caption{Overview of \methodname{} public methods employed for exporting articles' and authors' metadata .}
    \label{tab:pybiblionet_functions_exporting}
    \end{table}

    \item \textbf{Data Analysis}
    Table~\ref{tab:pybiblionet_functions_dataanalysis} presents the functions designed to support data analysis. These methods enable users to perform both temporal and ranking analyses, facilitating the interpretation of publication dynamics, topic evolution, and scholarly impact. In particular, temporal data analysis functions such as \texttt{plot\_article\_trends}, \texttt{plot\_topic\_trends}, and \texttt{plot\_keyword\_trends} allow the visualization of longitudinal patterns in article production, thematic shifts, and keyword prominence. Ranking functions like \texttt{plot\_top\_authors} and \texttt{plot\_top\_keywords\_from\_abstracts} support the identification of influential authors and the most frequent terms in article abstracts, optionally accounting for citation counts and the role of peripheral articles.

    \begin{table}[h!]
    \centering
    \footnotesize
    \renewcommand{\arraystretch}{1.3} 
    \begin{tabularx}{\linewidth}{|>{\raggedright\arraybackslash}p{0.25\linewidth}|>{\raggedright\arraybackslash}p{0.2\linewidth}|>{\raggedright\arraybackslash}X|}
    \hline
    \textbf{Method} & \textbf{Task} & \textbf{Description} \\ 
    \hline
    \texttt{plot\_article\_trends} & Temporal data analysis & Plots a stacked area chart that visualizes the number of root set and non-root set articles over time, aggregated by a specified interval. \\ \hline
    \texttt{plot\_topic\_trends} & Temporal data analysis & Plots a grouped stacked bar chart for aggregated data based on topics. \\ \hline
    \texttt{plot\_keyword\_trends} & Temporal data analysis & Plots a line chart showing the trends of top keywords over time. \\ \hline
    \texttt{plot\_top\_authors} & Ranking analysis & Plots a stacked bar chart showing the top authors and their articles or citations split by topics. \\ \hline
    \texttt{plot\_top\_keywords} \texttt{\_from\_abstracts} & Ranking analysis & Plots the most frequent keywords extracted from abstracts of articles, including base set when \texttt{show\_periphery=True}. \\ \hline
    
    \end{tabularx}
    \caption{Overview of \methodname{} public methods employed for data analysis.}
    \label{tab:pybiblionet_functions_dataanalysis}
    \end{table}

\end{itemize}

\subsection{Data Analytics}
Prior to the analysis of the citation network, it might be useful to obtain descriptive statistics of the corpus of articles gathered by the data collection module. This can be achieved through a set of functions contained in the \texttt{bibliometric\_analysis.charts} module. This module includes functions to:

\begin{itemize}
    \item {Temporal Analysis}: Three public methods (i.e. \texttt{plot\_article\_trends}, \texttt{plot\_topic\_trends}, and \texttt{plot\_keyword\_trends}) allow for temporal trend analysis of the number of published articles, the topics covered, and the associated keywords.

    \item {Ranking Analysis}: The public methods \texttt{plot\_top\_authors} and \texttt{plot\_top\_keywords\_from\_abstracts} allow for the evaluation of the most prolific (or most cited) authors and the most relevant keywords extracted from the abstracts of all retrieved articles, respectively.

\end{itemize}
The topic is one of the entities in the OpenAlex database and can be analyzed in terms of ``field'' or ``domain'' where domain is the highest level in the "domain, field, subfield, topic" system.
For keyword extraction, we employed KeyBERT~\cite{grootendorst2020keybert} to identify the most relevant terms. KeyBERT utilizes BERT embeddings to extract the most pertinent words or phrases from a given text. This approach enables the detection of key terms within abstracts over time, offering insights into their temporal evolution.

\subsection{Network-based Bibliometrics}\label{subsec:code}

The core of the bibliometric\_analysis module deals with the analysis of the citation and co-authorship graphs in multiple ways. It supports the examination of structural properties, the identification of influential articles and authors, allowing users to extract centrality metrics, and performing clustering.

\begin{table}[h!]
\centering
\footnotesize
\renewcommand{\arraystretch}{1.3} 
\begin{tabularx}{\linewidth}{>{\raggedright\arraybackslash}p{0.25\linewidth}>{\raggedright\arraybackslash}p{0.2\linewidth}>{\raggedright\arraybackslash}X}
\hline
\textbf{Method} & \textbf{Task} & \textbf{Description} \\ 
\hline
\texttt{extract\_metrics\_to\_csv} & Extracting Network Metrics & Computes centrality metrics (e.g., betweenness, closeness, PageRank) for nodes in a citation network and exports the results to a CSV file. \\ \hline
\texttt{show\_graph\_statistics} & Visualizing Network Metrics & Displays the distribution of centrality metrics (e.g., betweenness, closeness, PageRank) through histograms, presents the correlation heatmap of centrality measures, and provides a table with statistics of the citation graph. \\  \hline
\texttt{cluster\_graph} & Clustering Articles & Assigns clusters to articles in the network using algorithms like Louvain, Girvan-Newman, and Infomap, enabling the exploration of thematic groupings. \\ \hline
\texttt{extract\_clusters\_to\_csv} & Exporting Clustering Results & Exports cluster assignments and related node information to a CSV file for further analysis. \\ \hline
\texttt{show\_clustered\_graph} & Visualizing Clusters & Visualizes cluster structures in the bibliometric graph, using pie charts to depict entity distributions and providing customization options for presentation. \\ \hline
\texttt{show\_cluster\_statistics} & Visualization of Cluster Sizes & Generates bar charts to illustrate the distribution of cluster sizes within the network. \\ \hline

\end{tabularx}
\caption{Overview of \methodname{} public methods for bibliometric analysis and clustering.}
\label{tab:pybiblionet_functions_bibliometric}
\end{table}

Table~\ref{tab:pybiblionet_functions_bibliometric} provides an overview of the available functions, which include capabilities to:

\begin{itemize}
    \item \textbf{extract network metrics}: once the citation graph is built, the \texttt{extract\_metrics\_to\_csv} function calculates various centrality metrics for nodes (i.e., articles or authors) within a citation or a co-authorship network, and it saves these metrics along with other selected fields in a CSV file. The supported metrics currently include betweenness centrality, closeness centrality, eigenvector centrality, pagerank, degree, in-degree and out-degree (Table \ref{tab:centralities}).  These metrics are essential for identifying influential papers within a citation network. For example, high betweenness centrality may indicate a paper that connects different clusters of research, while PageRank can help identify papers that are frequently cited by influential works.

    \item \textbf{visualize network statistics}: the method show\_graph\_statistics generates customized graphs that facilitate a more detailed exploration of network statistics. The visualizations include a summary of overall network statistics, histograms displaying the distribution of each centrality metric, and a correlation matrix illustrating relationships among centrality metrics.

    \item \textbf{perform community detection}: the \texttt{cluster\_graph} function assigns clusters to articles within the citation network, or to authors within the co-authorship network, using one of the community detection algorithms listed in Table \ref{tab:community_detection}. The function returns a graph where each node is annotated with a cluster identifier, facilitating the exploration of research domains or thematic groupings within the dataset. While supported algorithms currently include Louvain, Girvan-Newman, Infomap, Spectral Clustering, and Stochastic Block Models, other algorithms can easily be included according to the necessities discussed in Sec.~\ref{subsec:comdet}. Node-cluster associations can also be exported to CSV for further analysis through the method \texttt{extract\_clusters\_to\_csv}.

    \item \textbf{visualize communities}: a key feature of \methodname{} is the visualization of the community structure of the bibliometric graph. The visualization revolves around the \texttt{show\_clustered\_graph} function. It provides an intuitive representation of clusters of articles or authors in a bibliometric network, utilizing pie charts to depict the distribution of topics (for articles) or countries (for authors) within each cluster. This function also allows for various customization options to tailor the visual output to specific needs, such as adjusting the color and size of nodes and topics, as well as the thickness of pie charts.

\end{itemize}

\section{Use Case: Application of \textit{\methodname{}} on the Query ``15 Minute City''}\label{sec:usecase}

In this section, we show the use of the \textit{\methodname{}} library for conducting a bibliometric analysis on the query ``15-minute city'', an urban planning concept that promotes access to essential services within a 15-minute walk or bike ride from home \cite{moreno2021introducing}. This topic gained increasing interest in recent years due to its association with urban planning and sustainable development concepts. Also, it has been studied by scholars from different disciplines and perspectives, so that we can more easily showcase how effective a clusters analysis can be.

The analysis comprises different phases: the retrieval of relevant articles and the construction of a citation network, followed by the extraction of key bibliometric metrics and the clustering of the citation and co-authorship graphs. Below, we describe the execution of these steps in detail.

\subsection{Retrieving Articles and Constructing the Citation Graph}
\label{sec:retrieving}

In the following example, we perform the first step of a typical application of \methodname{} by retrieving articles based on the bibliometric query and by creating a citation network using the methods described in Table \ref{tab:pybiblionet_functions_retriving}. The following code snippet shows the process:

\begin{lstlisting}
import json

from pybiblionet.openalex.core import string_generator_from_lite_regex, retrieve_articles,create_citation_graph,create_coauthorship_graph

if __name__ == "__main__" :


    queries = string_generator_from_lite_regex("(15)( )(minute|min)( )(city)")

    mail = "your_email@example.com"
    from_publication_date = "2019-01-01"
    to_publication_date = None
    periphery = True

    json_file_path = retrieve_articles(queries, mail, from_publication_date, to_publication_date)
    #the articles are saved in a json file inside the query_result folder
    articles = json.load(open(json_file_path))
    G_citation=create_citation_graph(articles,"citation_graph.GML",periphery=True)
    print(G_citation.number_of_nodes(),G_citation.number_of_edges())
    G_coauthorship=create_coauthorship_graph(articles,"coauthorship_graph.GML",periphery=True)
    print(G_coauthorship.number_of_nodes(),G_coauthorship.number_of_edges())

\end{lstlisting}

This script uses the OpenAlex API to search for articles based on the query ``15 minute/min city.'' As we can see, we deployed some of the methods described in the previous section, that conveniently wrap all the operations involved in parsing the query, calling the appropriate OpenAlex API functions, retrieving and parsing the correct metadata upon which we will build the citation graph.

Indeed, the query is generated by the \texttt{string\_generator\_from\_lite\_regex} function, allowing for flexibility in how the query terms are combined. In this case, we specified the date range starting from January 1st, 2019. The \texttt{retrieve\_articles} function is then executed with these parameters to collect the relevant articles, including those in the base set, and store them in a JSON file within the \texttt{query\_result} folder.

Optionally, the retrieved articles can be exported as a CSV file by specifying the desired fields (e.g., \texttt{export\_path = ``15minute.csv''}). The retrieved articles are then used to construct a citation graph using the \texttt{create\_citation\_graph} function. In this example, base set articles are included in the graph construction. The resulting graph is saved in GML format, which is well-suited for further analysis in the next step.
As a standard and widely supported format, GML can also be used as input for external network analysis tools such as Gephi, thereby facilitating seamless integration into existing analytical tools.

\subsection{Metadata Export}

The library also supports exporting collected data in CSV format, making it more accessible for users who may find the JSON output of the \texttt{retrieve\_articles} method less convenient to explore. This functionality allows the export of selected metadata related to articles, authors, institutions, and venues. For institutions and venues, occurrence counts within the root set and base set are also included. These export capabilities are implemented in four methods within the core module, summarized in Table \ref{tab:pybiblionet_functions_exporting} (more details about these methods in Appendix \ref{appendixB}). Users can customize the output by selecting specific fields and specifying the filename. To enhance interoperability with existing libraries and tools for bibliometric analysis, an additional feature allows exporting articles in Scopus-compatible formats, which are commonly used as input by tools such as Bibliometrix \cite{aria2017bibliometrix}, Scientopy \cite{Scientopy2019}, and VOSviewer \cite{van2010software}. Although not all data fields available in Scopus are accessible through OpenAlex, we mapped all compatible elements to maximize the amount of usable information and support seamless integration with external tools.
The following code snippet demonstrates how to export article metadata either in a user-defined CSV format—by selecting specific fields to include—or in a Scopus-like CSV format, which can then be used as input for the Python library pybibx \cite{pereira2023pybibx}:

\begin{lstlisting}
from pybiblionet.openalex.core import export_articles_to_csv, export_authors_to_csv
from pybibx.base import pbx_probe

if __name__ == "__main__" :


    json_file_path= "query_results/query_result_1ea020320b230de5a973a39682eaa53dce89a9bb026b441a5f825232.json"

    export_articles_to_csv(json_file_path,
                           fields_to_export=["venue","type","id", "doi", "title", "publication_date","authorships_display_name","language"],
                           export_path="15minute_articles.csv"
                           )

    export_articles_to_scopus(json_file_path,
        export_path="scopus/scopus.csv",
        include_periphery=True
    )

    filename = "scopus/myscopus.csv"
    bibfile = pbx_probe(file_bib=filename,db="scopus",del_duplicated=True)

    report= bibfile.eda_bib()
\end{lstlisting}

\subsection{Data Analysis}

\begin{figure}[h!]
    \centering

    \begin{subfigure}[t]{0.48\textwidth}
      \centering
      \captionsetup{justification=centering} 
            \includegraphics[width=\textwidth]{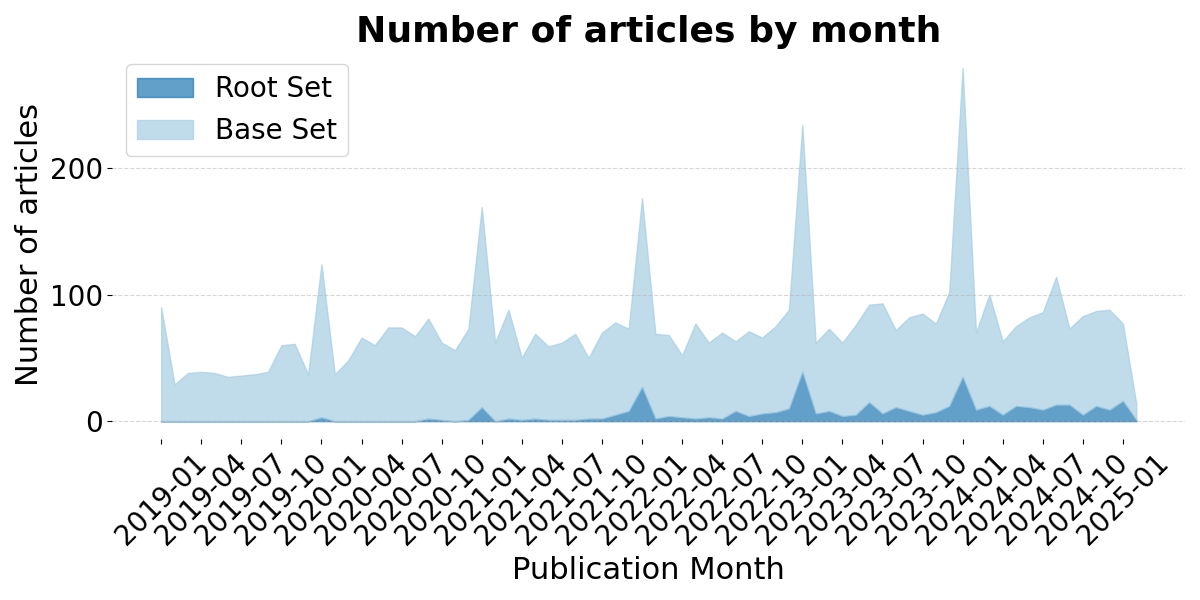}
            \caption{}
            \label{fig:data_analisis_a}
        \end{subfigure}
    \begin{subfigure}[t]{0.48\textwidth}
        \centering
        \captionsetup{justification=centering} 
        \includegraphics[width=\textwidth]{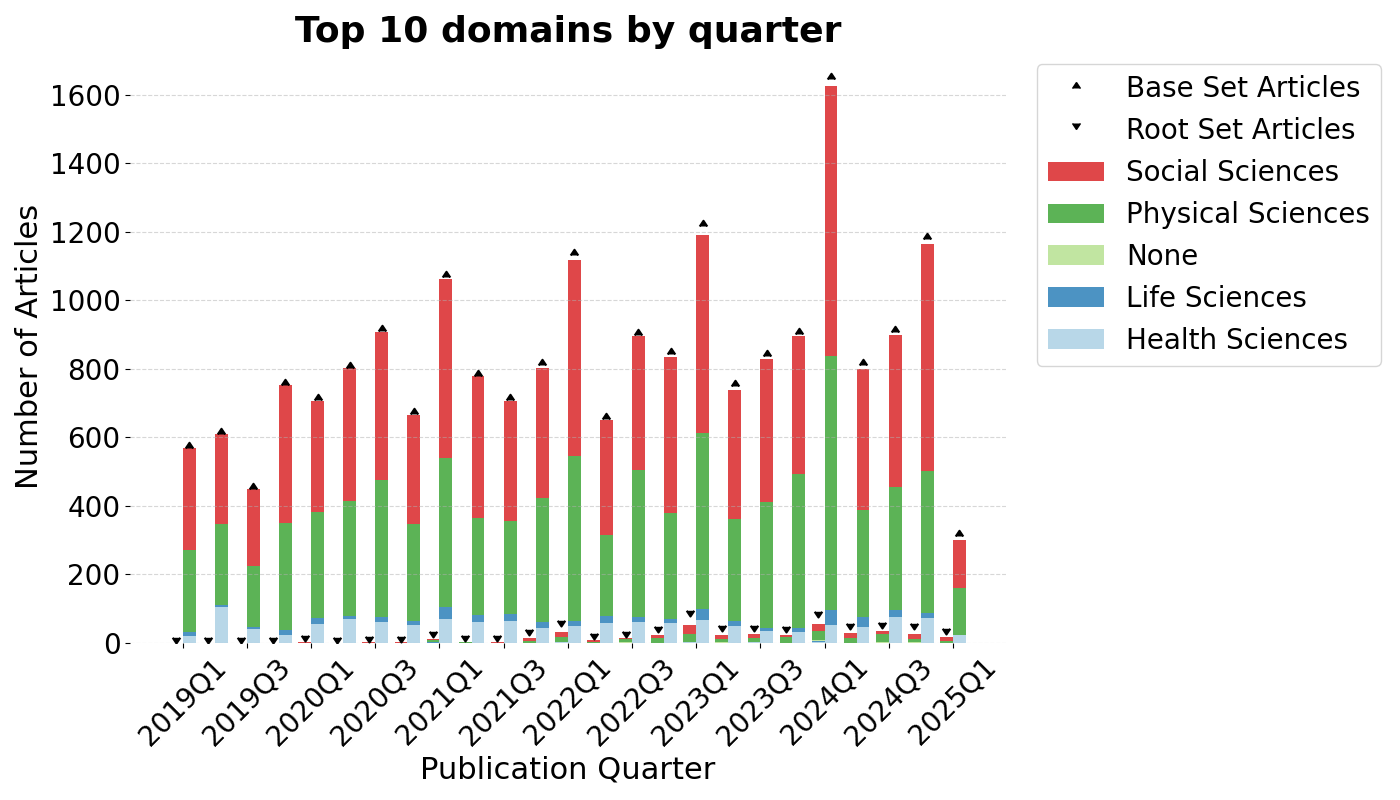}  
        \caption{}
        \label{fig:data_analisis_b}
    \end{subfigure}
    \begin{subfigure}[t]{0.48\textwidth}
        \centering
        \captionsetup{justification=centering}         \includegraphics[width=\textwidth]{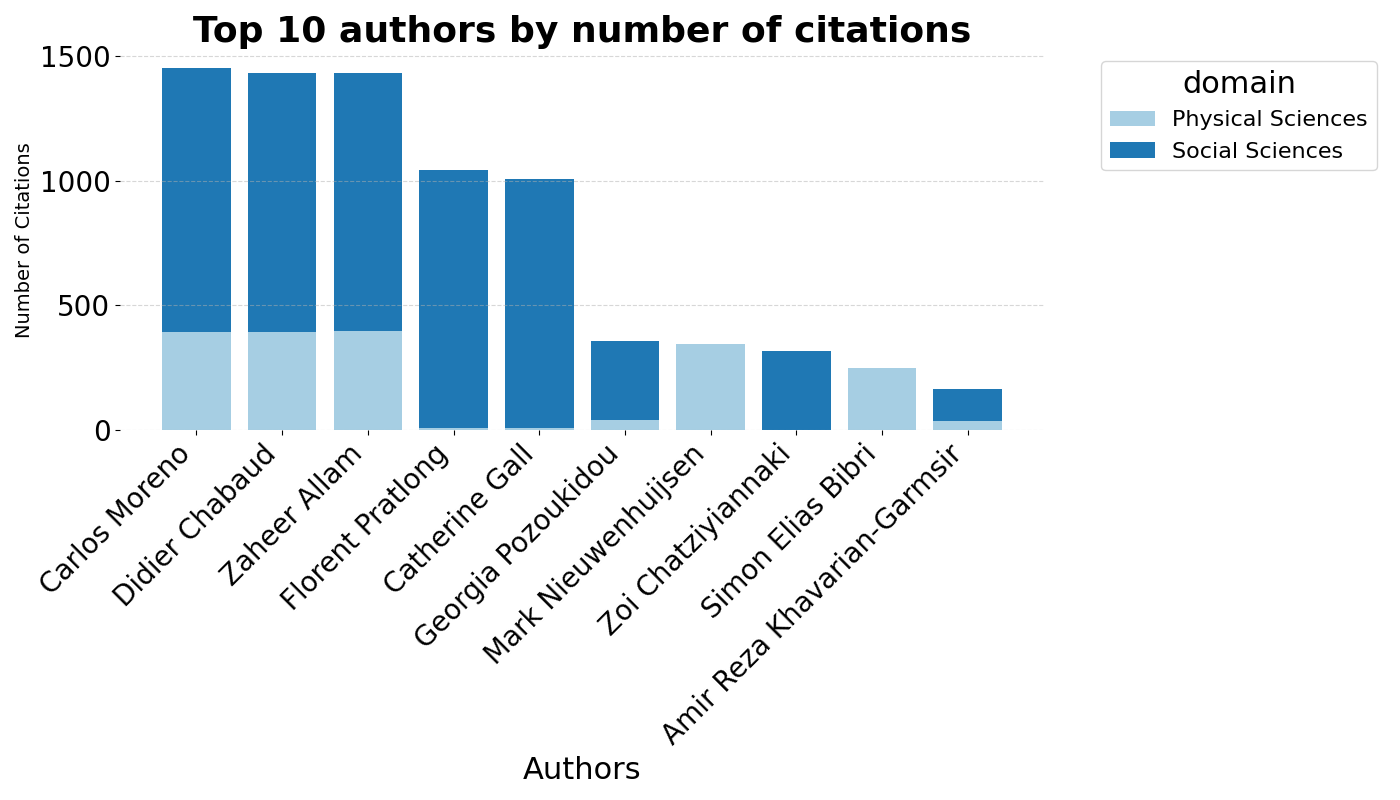}  
        \caption{}
        \label{fig:data_analisis_c}

    \end{subfigure}
        \begin{subfigure}[t]{0.48\textwidth}
        \centering
        \captionsetup{justification=centering}         \includegraphics[width=\textwidth]{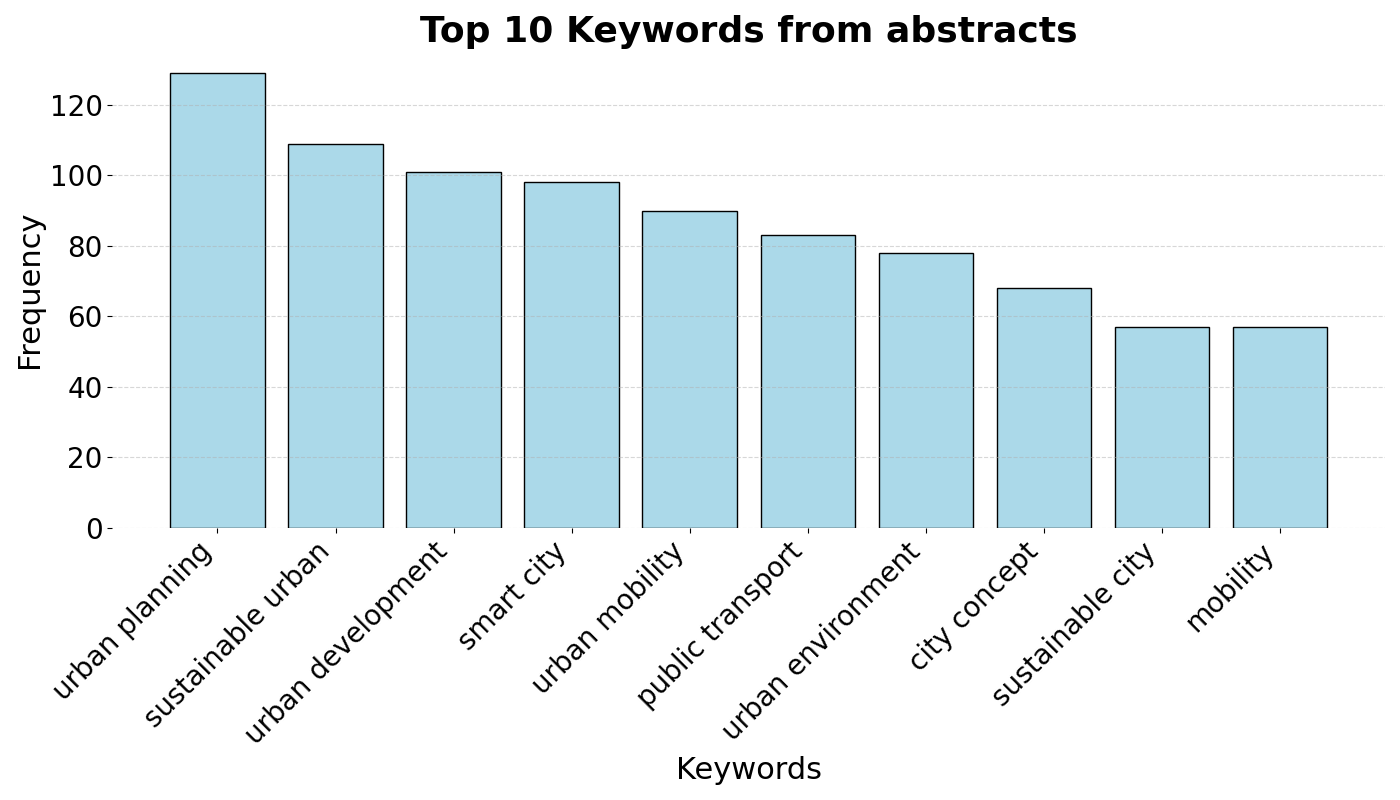}  
        \caption{}
        \label{fig:data_analisis_d}
    \end{subfigure}
    \begin{subfigure}[t]{0.48\textwidth}
        \centering
        \captionsetup{justification=centering}         \includegraphics[width=\textwidth]{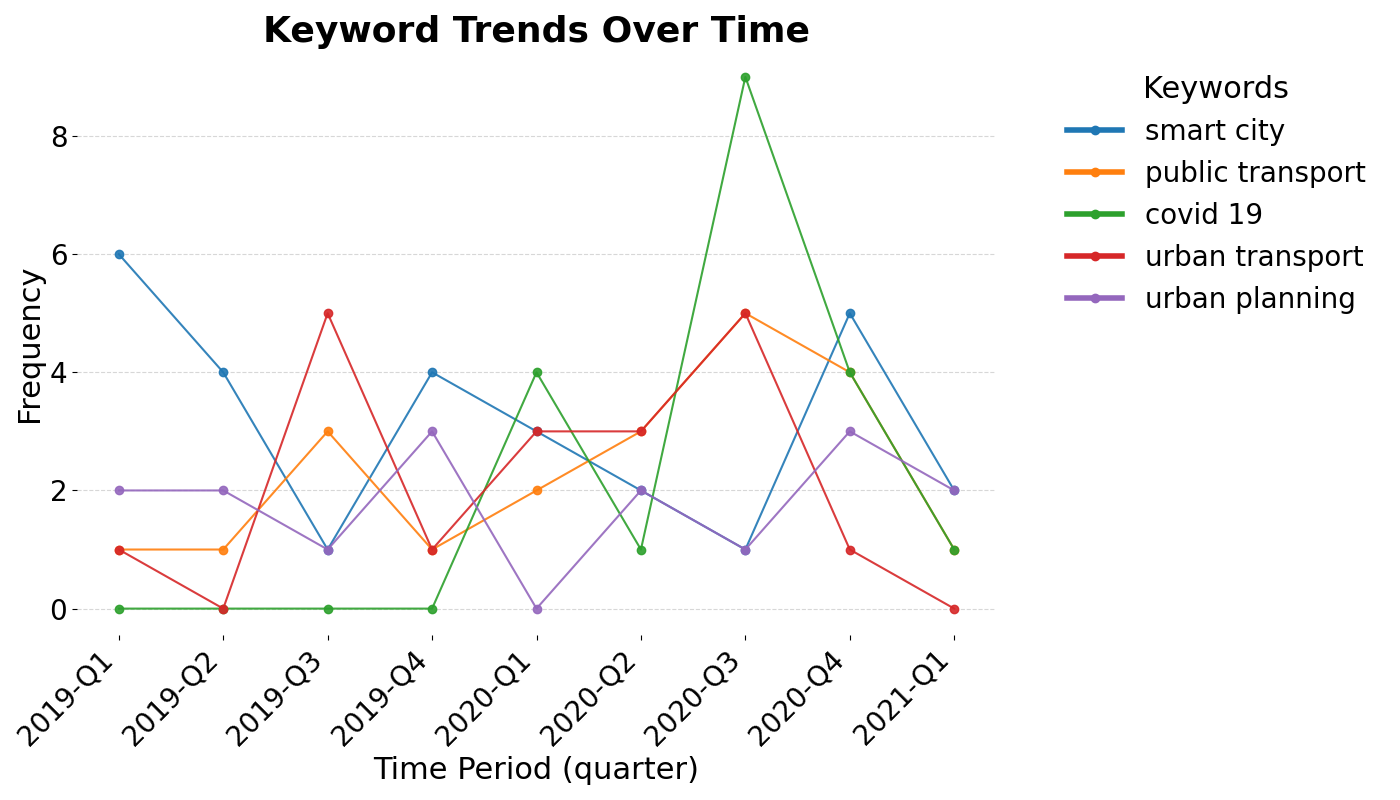} 
        \caption{}
        \label{fig:data_analisis_e}

    \end{subfigure}
    \caption{Temporal analysis of published articles and research trends. (a) Number of articles published per quarter. (b) Top concepts appearing in articles by month. (c) Top 10 authors ranked by the number of published articles. (d) Most frequent keywords extracted from abstracts. (e) Trends of top keywords over time.}
    \label{fig:data_analisis}
\end{figure}

Table \ref{tab:pybiblionet_functions_dataanalysis} presents the methods provided by the library for analyzing the temporal distribution of articles recovered by the public method \texttt{retrieve\_articles} (more details in Appendix \ref{appendixB}). These methods enable various types of visualizations, including stacked area charts illustrating the distribution of root set and base set articles over specified intervals, grouped stacked bar charts for aggregated topic-based data, stacked bar charts showing top authors and the distribution of their primary topic contributions, keyword frequency analysis from abstracts, and keyword trend analysis over time.
The code snippets presented in the algorithm demonstrate that articles can be loaded from the same file generated using \texttt{retrieve\_articles} in Section \ref{sec:retrieving} and that generate the charts shown in Figure \ref{fig:data_analisis}. The methods that generate the graphs are customizable in terms of style, allowing for personalized visualization by adjusting different levels of temporal aggregation or limiting the number of displayed entries.

\begin{lstlisting}
from pybiblionet.bibliometric_analysis.charts import plot_topic_trends, plot_article_trends, \
    plot_top_authors, plot_top_keywords_from_abstracts, plot_keyword_trends
from datetime import datetime
import matplotlib.pyplot as plt
from matplotlib import colors
import numpy as np
import json
if __name__ == "__main__" :


    json_file_path= "query_results/query_result_1ea020320b230de5a973a39682eaa53dce89a9bb026b441a5f825232.json"
    articles = json.load(open(json_file_path))

    plot_article_trends(articles,
                        date_from=datetime(2020, 1, 1),
                        interval="quarter",  # Change to "month" or "quarter" or "year" as needed
                        date_to=datetime(2023, 12, 31),
                        num_ticks=20)
    plot_topic_trends(
        articles,
        top_n_colors= [colors.to_hex(c) for c in  plt.get_cmap('tab10')(np.linspace(0, 1, 10))],
        show_core=True,
        show_periphery=True,
        interval="month", # Change to "quarter" or "year"
        date_from=datetime(2021, 9, 1),
        date_to=datetime(2022, 12, 31),
        top_n=10,
        )
    plot_top_authors(
        articles,
        date_from=datetime(2020, 1, 1),
        date_to=datetime(2022, 12, 31),
        num_authors=10,
        by_citations=True,
        show_periphery=False,
        n_colors=["#1f77b4", "#ff7f0e", "#2ca02c", "#d62728", "#9467bd"]
    )

    plot_top_keywords_from_abstracts(articles)


    plot_keyword_trends(
        articles=articles,
        date_from=datetime(2020, 1, 1),
        date_to=datetime(2023, 1, 1),
        show_core=True,
        show_periphery=True,
        top_n=10,
        ngram_range=(1, 2),
        interval="quarter"
    )

\end{lstlisting}

Figure \ref{fig:data_analisis_a} presents the distribution of root set and base set articles through a stacked area chart. Root set articles are those that explicitly include the search string "15 minutes" in their title or abstract, while base set articles do not. The first article identified using this search criterion \footnote{it is important to note that \texttt{retrieve\_articles} was also configured with a date filter, restricting results to publications after January 1, 2019. Consequently, no root set articles published before this date are included in the retrieved dataset} dates back to January 2020. Although the ``15-minute city'' paradigm was initially proposed by Carlos Moreno in 2016, it gained significant popularity only in 2020 when the Mayor of Paris, Anne Hidalgo, advocated for its adoption during her re-election campaign.
Indeed, from 2020 onwards, the number of root set articles has increased, alongside the growth in base set articles. Notably, peaks in publications are observed in January, aligning with well-known trends in scientific publishing. One possible explanation is that gratuity fees reset at the beginning of the new year, or that citations are counted starting from the following year. As a result, articles published early in the year may have a longer citation lifespan, potentially increasing their overall impact. This trend suggests that articles on this topic not only cite earlier works but also frequently reference contemporary research.
Figures \ref{fig:data_analisis_b} and \ref{fig:data_analisis_c} provide further insights. Carlos Moreno emerges as a highly cited author among root set articles, often in collaboration with co-authors such as Allam and Chabaud, and the main topic domains are Social Science and Physical Science. This finding is unsurprising given the fact that Moreno was the first that proposed the 15-minute paradigm. Figure \ref{fig:data_analisis_b} further reveals that base set articles include a substantial number of publications focusing on COVID-19 epidemiological studies. These articles began appearing shortly after the onset of the pandemic and are frequently cited by root set articles. This correlation likely stems from the fact that the 15-minute city concept aligns well with pandemic-era urban planning needs, facilitating localized access to essential services and supporting lockdown measures.
Beyond pandemic-related studies, the majority of topics covered in base set articles pertain to mobility, urban planning, and smart cities, highlighting the broader relevance of the 15-minute city paradigm within these research domains.

Figures \ref{fig:data_analisis_d} and \ref{fig:data_analisis_e} illustrate the results of keyword extraction using KeyBERT\footnote{KeyBERT may return morphologically similar keywords due to the lack of lemmatization, treating singular and plural forms as distinct despite their semantic equivalence (e.g. \textit{smart city} and \textit{smart cities}).}. Specifically, Figure  \ref{fig:data_analisis_d} presents a simple bar chart displaying the most frequent keywords in root set articles, while Figure  \ref{fig:data_analisis_e} additionally depicts temporal trends including base set articles. Although the overall frequency of keywords increases over time, certain terms, including ``city covid'' and ``city concept'', exhibit a more pronounced growth. This initial phase of analysis provides valuable insights into the evolution of discourse on this topic over time, the most involved authors, and related subtopics

\subsection{Network-based Bibliometrics}

Once we retrieve all the relevant information pertaining the OpenAlex query, upon which we built a citation a co-authorship graph, we can exploit some of the methods implemented in \methodname{}, and described in Sec.~\ref{subsec:code}, to perform the bibliometric analysis, the community detection and the visualization of the results (a summary of the method is provided in \ref{tab:pybiblionet_functions_bibliometric}).

\begin{lstlisting}
import networkx as nx
import numpy as np
import matplotlib.pyplot as plt
from pybiblionet.bibliometric_analysis.core import extract_metrics_to_csv, cluster_graph, extract_clusters_to_csv
from pybiblionet.bibliometric_analysis.charts_network import show_clustered_graph, show_cluster_statistics,show_graph_statistics
from matplotlib import colors


if __name__ == "__main__":
    print("Reading graph...")
    
    json_file_path="query_results/query_result_1ea020320b230de5a973a39682eaa53dce89a9bb026b441a5f825232.json"
    
    network_file_name= "15minute_citation_graph.GML"

    G = nx.read_gml(network_file_name)
   

    print("Extracting metrics...")
    metrics = ["betweenness_centrality", "closeness_centrality", "page_rank", "in_degree", "out_degree"]
    fields = ["id", "title"]
    csv_file_path = "metrics_and_fields.csv"

    
    extract_metrics_to_csv(G, metrics, fields, csv_file_path)

    show_graph_statistics(G,csv_file_path)


    print("\nClustering the graph...")
    clustered_graph = cluster_graph(G, algorithm='louvain',)
    csv_file_path = "cluster_and_fields.csv"
    extract_clusters_to_csv(clustered_graph, fields, csv_file_path)


    print("Visualizing clusters...")
    show_clustered_graph(clustered_graph, image_size=(800, 800),
                         n_clusters=5,
                         topics_level="field",
                         )

    show_cluster_statistics(csv_file_path, image_size=(800, 800),
                 n_clusters=5)
\end{lstlisting}

\begin{figure}[h!]
    \centering
    \begin{subfigure}[t]{0.48\textwidth}
        \centering
        \captionsetup{justification=centering} 
        \includegraphics[width=\textwidth]{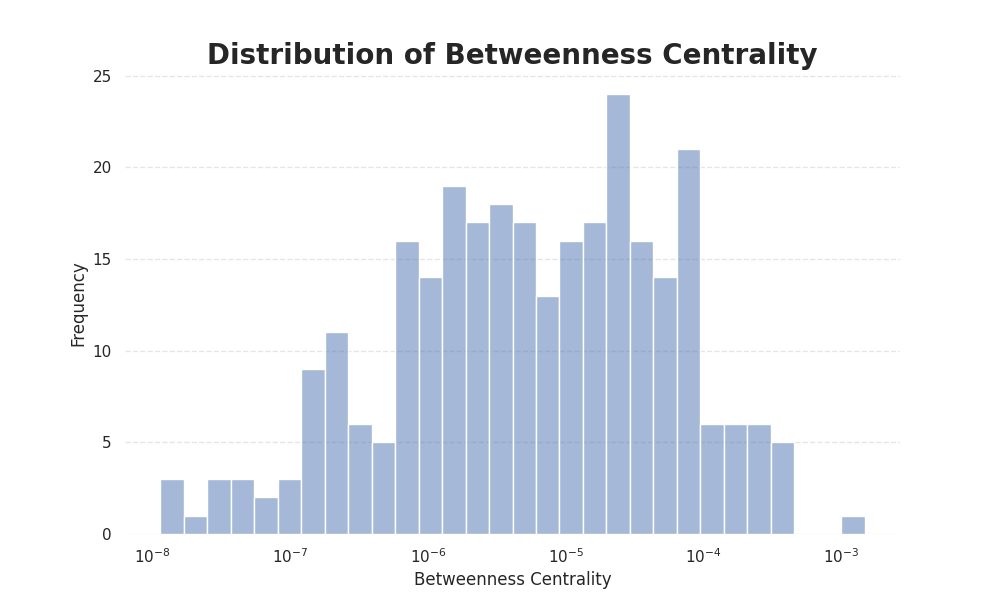}  
        \caption{}
        \label{fig:table_and_heatmap_a}
    \end{subfigure}
    \begin{subfigure}[t]{0.48\textwidth}
        \centering
        \captionsetup{justification=centering} 
        \includegraphics[width=\textwidth]{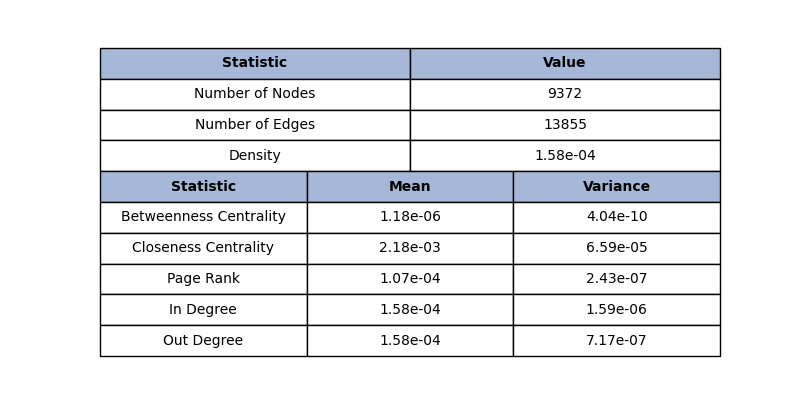} 
        \caption{}
        \label{fig:table_and_heatmap_b}
    \end{subfigure}
    \begin{subfigure}[t]{0.48\textwidth}
        \centering
        \captionsetup{justification=centering} 
        \includegraphics[width=\textwidth]{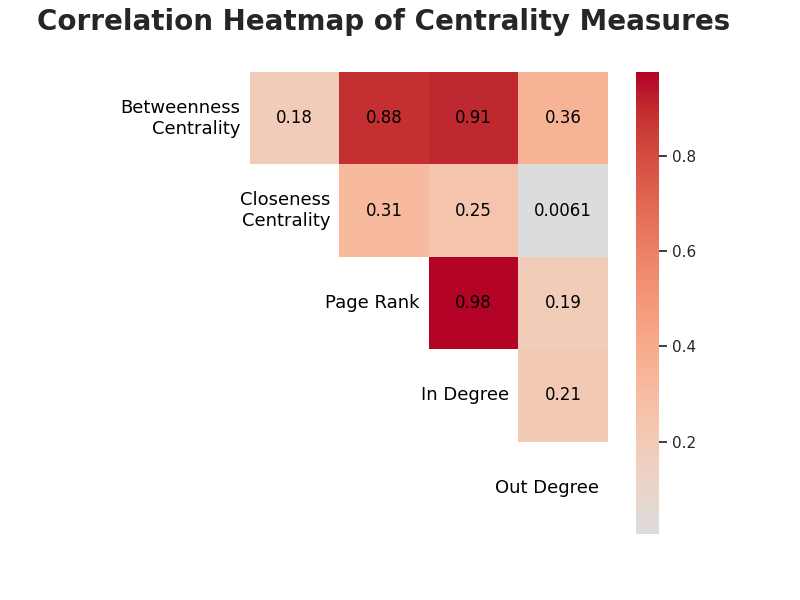}  
        \caption{}
        \label{fig:table_and_heatmap_c}
    \end{subfigure}
    \caption{(a) It presents the distribution of the sampled metrics, with betweenness centrality shown as an example in this image. (b) Statistics of the citation graph obtained by querying "15 minutes city" using \methodname{}. In this graph, nodes are scientific papers that are linked by an edge if there is an outgoing or incoming citation between them. (c) Correlation heatmap of centrality measures.}
    \label{fig:table_and_heatmap}
\end{figure}

In the snippet of code, we first load the citation graph from the GML file exported in Section \ref{sec:retrieving} and then extract various centrality measures, such as \textit{betweenness centrality}, \textit{closeness centrality}, and \textit{PageRank} (calling the function \texttt{extract\_metrics\_to\_csv}).These metrics are stored in a CSV file along with key information about the articles (e.g., title, abstract, authors), providing insight into the importance of different articles within the citation network. The function \texttt{show\_graph\_statistics} allows for the visualization of general network statistics, the distribution of each metric, and the correlation heatmap of metrics (Figures ~\ref{fig:table_and_heatmap}).
Although the same operation can be performed for the co-authorship network, it is important to note that this network is undirected. Consequently, in-degree and out-degree metrics are not applicable. Similarly, the types of fields that can be exported for each author differ from those available for articles.
The export of metadata from articles associated with network measures allows for the identification of the most relevant articles for each metric, even without advanced computational expertise. By simply sorting the data using a program capable of reading CSV files (such as LibreOffice Calc), users can rank articles accordingly. In addition to identifying the most cited root set or base set articles, examining those with the highest eigenvector centrality, for instance, enables the detection of nodes with broader systemic impact, such as seminal papers that are frequently cited by other highly influential works.

\begin{figure}[h!]
    \centering
    \begin{subfigure}[b]{0.48\linewidth}
        \centering
        \captionsetup{justification=centering}
        \includegraphics[width=\linewidth]{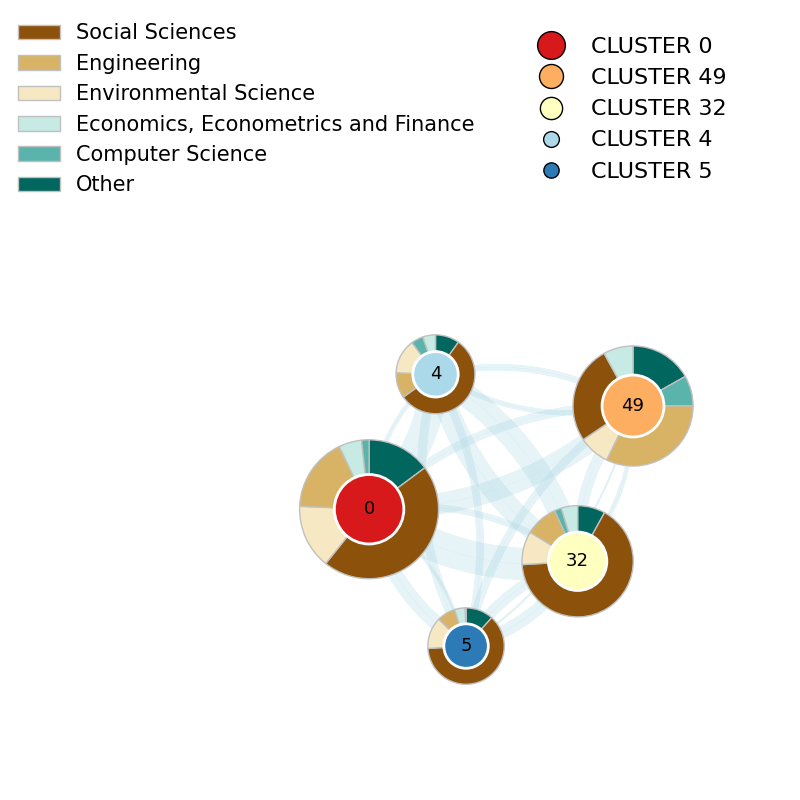}
        \caption{}
        \label{fig:clusters_a}
    \end{subfigure}
    \begin{subfigure}[b]{0.48\linewidth}
        \centering
        \captionsetup{justification=centering}
        \includegraphics[width=\linewidth]{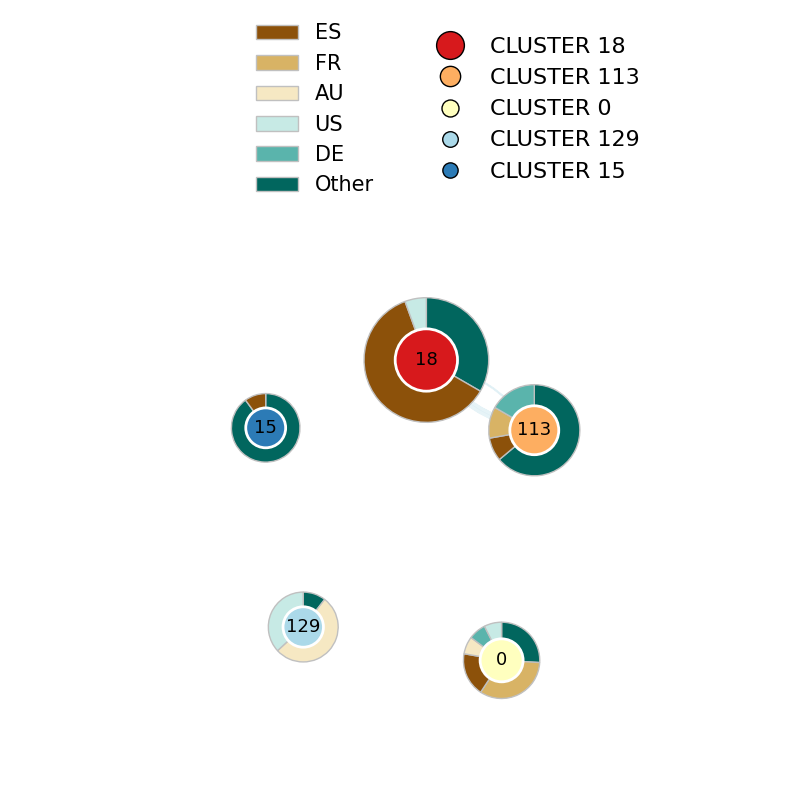}
        \caption{}
        \label{fig:clusters_b}
    \end{subfigure}
    \hfill
    \begin{subfigure}[b]{0.48\linewidth}
        \centering
        \captionsetup{justification=centering}
        \includegraphics[width=0.8\linewidth]{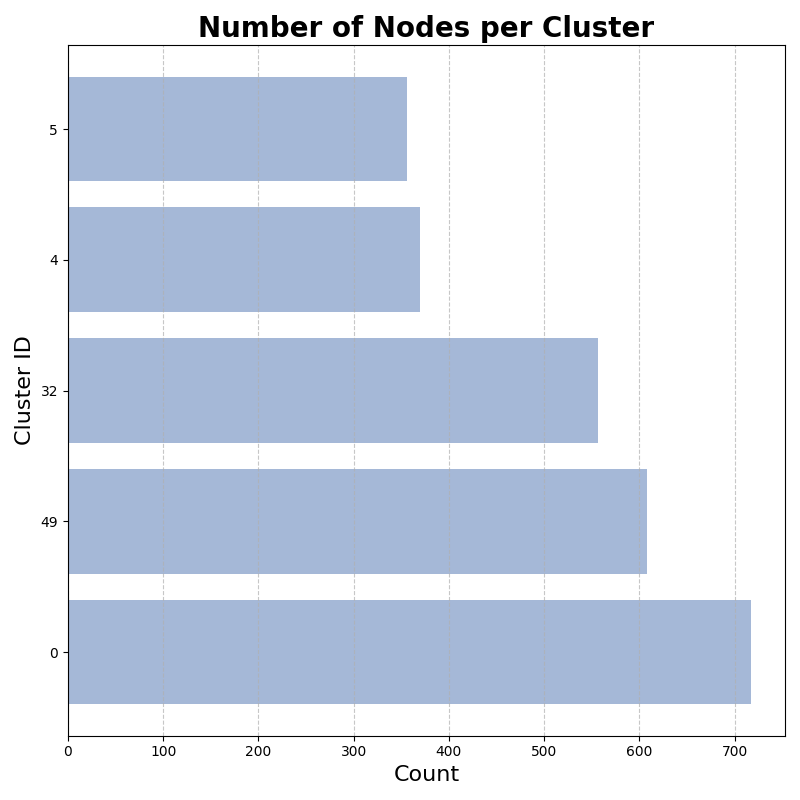}
        \caption{}
        \label{fig:clusters_hist_a}
    \end{subfigure}
        \begin{subfigure}[b]{0.48\linewidth}
        \centering
        \captionsetup{justification=centering}
        \includegraphics[width=0.8\linewidth]{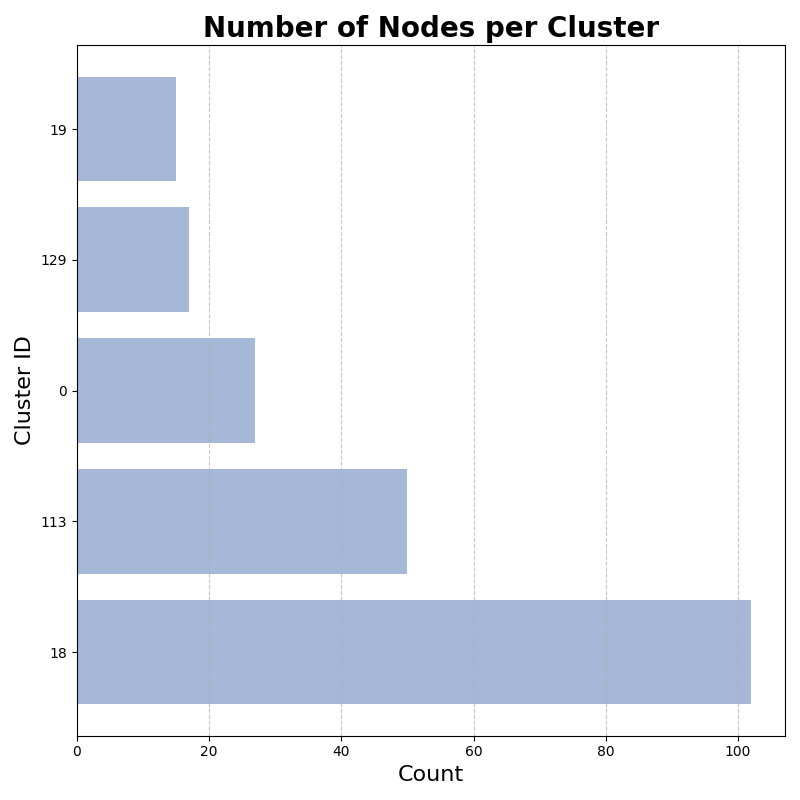}
        \caption{}
        \label{fig:clusters_hist_b}
    \end{subfigure}
    \caption{(a) and (b) Visualization of the clusters obtained through the function \textit{show\_clustered\_graph} respectively in the citation and co-authorship graphs. Each numbered cluster is surrounded by an indication of the main research areas of the papers, or country affiliation of the authors belonging to that particular cluster. (c) and (d) Visualization of the cluster sizes through the function \textit{show\_clusters\_chart} respectively in the citation and co-authorship graph.}
    \label{fig:clusters}
\end{figure}

We can then cluster the graphs exploiting the \textit{cluster\_graph} method, passing as an argument one of the implemented clustering techniques. For the purpose of this example the \textit{louvain} algorithm was chosen. The clustering results, along with article metadata, are exported into another CSV file for further examination. The clustered graph can be visualized using the \texttt{show\_cluster} function, which highlights the top 5 clusters by size. The output can be seen in Fig.~\ref{fig:clusters_a}. Each cluster is accompanied by a visual representation of its composition in terms of research areas (the most frequent 5 domains identified within the 5 largest clusters are displayed).
The distribution of key topics across the different clusters indicates that the subject ``15-minute city'' is multidisciplinary; however, there are clusters where certain main topics are more frequent. For example, one cluster exhibits a higher proportion of articles on ``Computer Science'' and ``Engineering'' compared to the others. The presence of articles on ``Environmental Science'' is not surprising, as the 15-minute city paradigm is viewed as a means to reduce private car use, thereby decreasing urban pollution within a framework centered on sustainability and well-being. Figure \ref{fig:clusters_hist_a} presents a bar chart depicting the size of the individual clusters. Figures \ref{fig:clusters_b} and \ref{fig:clusters_hist_b} display the same pair of graphs generated from the co-authorship network, the most frequent 5 countries within the 5 largest clusters are displayed. The key difference lies in the fact that the entities presented in the citation network correspond to the main topics of the articles in the cluster, whereas in the co-authorship network, the affiliation countries of the authors are shown.
The clusters in the co-authorship network appear more segregated, as there are no co-authorship relationships between the majority of the author clusters. Additionally, it becomes evident that collaboration between authors from different nationalities is limited to a few countries, predominantly Western, and specifically European. The large presence of authors with French affiliations can be explained by the fact that the model gained prominence in Paris.

\subsection{Results and Discussion}
The information retrieved by means of the different \methodname{} methods helps us in reconstructing an interesting bibliographical landscape around the topic of the 15-minutes city. By crawling data on the query "15 minutes/min city" starting from 2019, we are able to build a citation and a co-authorship graph whose basic statistics are synthesized in Fig.~\ref{fig:table_and_heatmap_a}. As seen, we retrieved a total of 9372 papers that, according to OpenAlex, satisfy the user query; by building the graph and analysing the connections among the papers, we are able to identify the most central papers and to explore more tightly connected clusters of research works, with the ultimate goal of narrowing down the large volume of results to identify the most relevant ones based on our objectives and needs. 

A simple analysis shows, on this very sparse network, a good level of correlation among the different centrality metrics, suggesting that overall there is good agreement in the importance of fundamental papers regardless of the specific definition of the centrality. However, once centrality is measured, it can be interesting to rank nodes and to pick a selection of most influential nodes according to the different definitions of centrality, to analyze specific variations and understand if there is only a very small number of influential papers playing more roles in the network, or if there is a diversity in terms of typology and research domains - and to quantify the extent of such variety.

This analysis can be further informed by the results of the community detection seen in Fig.\ref{fig:clusters_a}. The largest clusters identified by Louvain are composed of a variety of papers, which is already quite multidisciplinary, as we can see in the cluster composition shown in Fig.~\ref{fig:clusters_a}. Moreover, by inspecting the centrality-based node rankings in each community, we can identify the most prominent papers in each group. For this example, we only consider two well-known metrics, such as betweenness and closeness centrality. 
Tables~\ref{tab:betw} and~\ref{tab:closeness} show us that, beyond a couple of research works that are consistently prominent across different metrics, using more than one metric can already bring to the attention of the user different papers depending on the chosen metric. Many of these papers are in turn different from those that can be retrieved by simply looking at the list of the most cited research, shown in Tab.~\ref{tab:citations}. The composition of the communities is also non-trivial in terms of \textit{topics} and research areas, suggesting that we are dealing with a strongly multi-disciplinary query, and this must be kept into account when performing a bibliometric analysis on the topic.

\begin{table}[h]
\centering
\tiny
\resizebox{\linewidth}{!}{%
\begin{tabular}{p{0.05\linewidth} p{0.41\linewidth} p{0.29\linewidth} c c}
\toprule
\multicolumn{1}{c}{\textbf{Cluster}} & 
\multicolumn{1}{c}{\textbf{Title}} & 
\multicolumn{1}{c}{\textbf{Concepts}} & 
\multicolumn{1}{c}{\textbf{Rank}} &
\multicolumn{1}{c}{\textbf{Betweenness}} \\  \\ 
\multirow{3}{*}{\textbf{0}} &  Introducing the “15-Minute City”: Sustainability, Resilience and Place Identity in Future Post-Pandemic Cities \cite{moreno2021introducing}& Transportation (Urban Transport and Accessibility) & 
$1^\circ$ &
1.48e-03 \\
& \cellcolor{gray!15} 15-Minute City: Decomposing the New Urban Planning Eutopia \cite{pozoukidou202115minute}& \cellcolor{gray!15}Transportation (Urban Transport and Accessibility) & 
\cellcolor{gray!15}$9^\circ$ &
\cellcolor{gray!15}2.71e-04 \\
&  A Case Study of a 15-Minute City Concept in Singapore’s 2040 Land Transport Master Plan: 20-Minute Towns and a 45-Minute City \cite{manifesty2022case}& Transportation (Urban Transport and Accessibility) & 
$101^\circ$ &
1.94e-05 \\
\midrule
\multirow{3}{*}{\textbf{49}} &  Unpacking the ‘15-Minute City’ via 6G, IoT, and Digital Twins: Towards a New Narrative for Increasing Urban Efficiency, Resilience, and Sustainability \cite{allam2022unpacking}& Media Technology (Smart Cities and Technologies) & 
$8^\circ$&
3.02e-04 \\
& \cellcolor{gray!15} The Theoretical, Practical, and Technological Foundations of the 15-Minute City Model: Proximity and Its Environmental, Social and Economic Benefits for Sustainability \cite{allam2022theoretical}& \cellcolor{gray!15}Media Technology (Smart Cities and Technologies) &
\cellcolor{gray!15} $16^\circ$&
\cellcolor{gray!15} 1.65e-04 \\
&  Proximity-Based Planning and the “15-Minute City”: A Sustainable Model for the City of the Future \cite{Allam2023proximity}& Transportation (Urban Transport and Accessibility) & 
$17^\circ$&
1.62e-04 \\
\midrule
\multirow{3}{*}{\textbf{32}} &  A composite X-minute city cycling accessibility metric and its role in assessing spatial and socioeconomic inequalities – A case study in Utrecht, the Netherlands \cite{KNAP2023acomposite}& Transportation (Urban Transport and Accessibility) & 
$23^\circ$&
1.19e-04 \\
& \cellcolor{gray!15} Moving the 15-minute city beyond the urban core: The role of accessibility and public transport in the Netherlands \cite{POORTHUIS2023moving}& \cellcolor{gray!15}Transportation (Urban Transport and Accessibility) & 
\cellcolor{gray!15} $28^\circ$&
\cellcolor{gray!15} 8.77e-05 \\
&  Understanding the determinants of x-minute city policies: A review of the North American and Australian cities’ planning documents \cite{LU2023understanding} & Transportation (Urban Transport and Accessibility) &  
$29^\circ$&
8.75e-05 \\
\midrule
\multirow{3}{*}{\textbf{4}} &  Measuring compliance with the 15-minute city concept: State-of-the-art, major components and further requirements \cite{papadopoulos2023measuring}& Transportation (Urban Transport and Accessibility) & 
$4^\circ$& 
3.61e-04 \\
& \cellcolor{gray!15} The 15-minute city: Urban planning and design efforts toward creating sustainable neighborhoods \cite{KHAVARIANGARMSIR2023the15}& \cellcolor{gray!15}Transportation (Urban Transport and Accessibility) & 
\cellcolor{gray!15}$6^\circ$& 
\cellcolor{gray!15}3.36e-04 \\
&  The 15-minute city for all? – Measuring individual and temporal variations in walking accessibility \cite{WILLBERG2023the15}& Transportation (Urban Transport and Accessibility) & 
$10^\circ$& 
2.35e-04 \\
\midrule
\multirow{3}{*}{\textbf{5}} &  Urban Transition and the Return of Neighbourhood Planning. Questioning the Proximity Syndrome and the 15-Minute City \cite{marchigiani2022urban}& Management, Monitoring, Policy and Law (Urban Planning and Valuation) & 
$11^\circ$& 
2.24e-04 \\
& \cellcolor{gray!15} Barcelona under the 15-Minute City Lens: Mapping the Accessibility and Proximity Potential Based on Pedestrian Travel Times \cite{ferrer2022barcelona}& \cellcolor{gray!15}Transportation (Urban Transport and Accessibility) &
\cellcolor{gray!15}$12^\circ$& 
\cellcolor{gray!15}2.16e-04 \\
&  The Future of the City in the Name of Proximity: A New Perspective for the Urban Regeneration of Council Housing Suburbs in Italy after the Pandemic \cite{donofrio2022future}& Management, Monitoring, Policy and Law (Urban Planning and Valuation) &
$35^\circ$& 
7.86e-05 \\
\midrule

\bottomrule
\end{tabular}
}
\caption{Top Papers by Betweenness Centrality by Cluster}\label{tab:betw}
\end{table}

\begin{table}[h]
\centering
\tiny
\resizebox{\linewidth}{!}{%
\begin{tabular}{p{0.05\linewidth} p{0.41\linewidth} p{0.29\linewidth} c c}
\toprule
\multicolumn{1}{c}{\textbf{Cluster}} & 
\multicolumn{1}{c}{\textbf{Title}} & 
\multicolumn{1}{c}{\textbf{Concepts}} & 
\multicolumn{1}{c}{\textbf{Rank}} &
\multicolumn{1}{c}{\textbf{Closeness}} \\  \\ 
\midrule

\multirow{3}{*}{\textbf{0}} &  Introducing the “15-Minute City”: Sustainability, Resilience and Place Identity in Future Post-Pandemic Cities \cite{moreno2021introducing}& Transportation (Urban Transport and Accessibility) & $1^\circ$& 1.32e-01 \\
& \cellcolor{gray!15} Residents’ preferences for walkable neighbourhoods \cite{Brookfield2017Residents}& \cellcolor{gray!15}Transportation (Urban Transport and Accessibility) & \cellcolor{gray!15}$27^\circ$&\cellcolor{gray!15}8.03e-02 \\
&  Influence of mixed land-use on realizing the social capital \cite{NABIL2015Influence}& Sociology and Political Science (Urban, Neighborhood, and Segregation Studies) & $33^\circ$& 8.03e-02 \\
\midrule
\multirow{3}{*}{\textbf{49}} &  Redefining the Smart City: Culture, Metabolism and Governance \cite{allam2018redefining}& Media Technology (Smart Cities and Technologies) & $6^\circ$& 8.34e-02 \\
& \cellcolor{gray!15} Cities and the Digital Revolution \cite{allam2020cities}& \cellcolor{gray!15}Media Technology (ICT Impact and Policies) & \cellcolor{gray!15}$7^\circ$&\cellcolor{gray!15}8.29e-02 \\
&  On big data, artificial intelligence and smart cities \cite{ALLAM2019ONBIG} & Media Technology (Smart Cities and Technologies) & $8^\circ$&8.24e-02 \\
\midrule
\multirow{3}{*}{\textbf{32}} &  The 15-minute walkable neighborhoods: Measurement, social inequalities and implications for building healthy communities in urban China \cite{WENG2019the15} & Transportation (Urban Transport and Accessibility) & $2^\circ$&9.29e-02 \\
& \cellcolor{gray!15} Accessibility in Practice: 20-Minute City as a Sustainability Planning Goal \cite{capasso2019accessibility} & \cellcolor{gray!15}Transportation (Urban Transport and Accessibility) &\cellcolor{gray!15}$3^\circ$& \cellcolor{gray!15}8.65e-02 \\
&  The effect of COVID-19 and subsequent social distancing on travel behavior \cite{DEVOS2020theeffect}& Transportation (Urban Transport and Accessibility) & $19^\circ$&8.06e-02 \\
\midrule
\multirow{3}{*}{\textbf{4}} &  Negotiating Proximity in Sustainable Urban Planning: A Swedish Case \cite{gil2018negotiating}& Transportation (Urban Transport and Accessibility) & $78^\circ$&4.04e-02 \\
& \cellcolor{gray!15} Is accessibility an idea whose time has finally come? \cite{HANDY2020isaccessibility}& \cellcolor{gray!15}Transportation (Urban Transport and Accessibility) & \cellcolor{gray!15}$105^\circ$&\cellcolor{gray!15}2.74e-02 \\
&  The sustainable mobility paradigm \cite{BANISTER2008thesustainable} & Transportation (Urban Transport and Accessibility) & $113^\circ$&2.44e-02 \\
\midrule
\multirow{3}{*}{\textbf{5}} &  The Walkable city and the importance of the proximity environments for Barcelona’s everyday mobility \cite{MARQUET2015TheWalkable}& Transportation (Urban Transport and Accessibility) & $4^\circ$&8.50e-02 \\
& \cellcolor{gray!15} Transport and social exclusion in London \cite{CHURCH2000Transport}& \cellcolor{gray!15}Transportation (Urban Transport and Accessibility) &\cellcolor{gray!15}$73^\circ$& \cellcolor{gray!15}4.30e-02 \\
&  Barcelona under the 15-Minute City Lens: Mapping the Accessibility and Proximity Potential Based on Pedestrian Travel Times \cite{ferrer2022barcelona}& Transportation (Urban Transport and Accessibility) & $110^\circ$&2.49e-02 \\
\midrule

\bottomrule
\end{tabular}
}
\caption{Top Papers by Closeness Centrality by Cluster}\label{tab:closeness}
\end{table}

\begin{table}[h]
\centering
\tiny
\resizebox{\linewidth}{!}{%
\begin{tabular}{p{0.05\linewidth} p{0.41\linewidth} p{0.29\linewidth} c c}
\toprule
\multicolumn{1}{c}{\textbf{Cluster}} & 
\multicolumn{1}{c}{\textbf{Title}} & 
\multicolumn{1}{c}{\textbf{Concepts}} & 
\multicolumn{1}{c}{\textbf{Rank}} &
\multicolumn{1}{c}{\textbf{Citations}} \\  \\ 
\midrule

\multirow{3}{*}{\textbf{0}} &  Introducing the “15-Minute City”: Sustainability, Resilience and Place Identity in Future Post-Pandemic Cities \cite{moreno2021introducing}& Transportation (Urban Transport and Accessibility) & $1^\circ$&992 \\
& \cellcolor{gray!15} 15-Minute City: Decomposing the New Urban Planning Eutopia \cite{pozoukidou202115minute}& \cellcolor{gray!15}Transportation (Urban Transport and Accessibility) & \cellcolor{gray!15}$2^\circ$&\cellcolor{gray!15}318 \\
&  The 15-Minute City Quantified Using Mobility Data \cite{abbiasov2022the15}& Transportation (Human Mobility and Location-Based Analysis) & $59^\circ$&14 \\
\midrule
\multirow{3}{*}{\textbf{49}} &  The ‘15-Minute City’ concept can shape a net-zero urban future \cite{Allam2022the15} & Media Technology (Smart Cities and Technologies) & $7^\circ$&105 \\
& \cellcolor{gray!15} Unpacking the ‘15-Minute City’ via 6G, IoT, and Digital Twins: Towards a New Narrative for Increasing Urban Efficiency, Resilience, and Sustainability \cite{allam2022unpacking} & \cellcolor{gray!15}Media Technology (Smart Cities and Technologies) & \cellcolor{gray!15}$9^\circ$&\cellcolor{gray!15}87 \\
&  The Theoretical, Practical, and Technological Foundations of the 15-Minute City Model: Proximity and Its Environmental, Social and Economic Benefits for Sustainability \cite{allam2022theoretical} & Media Technology (Smart Cities and Technologies) & $17^\circ$&56 \\
\midrule
\multirow{3}{*}{\textbf{32}} &  Is the 15-minute city within reach? Evaluating walking and cycling accessibility to grocery stores in Vancouver \cite{HOSFORD2022isthe15}& Transportation (Urban Transport and Accessibility) & $15^\circ$&57 \\
& \cellcolor{gray!15} A composite X-minute city cycling accessibility metric and its role in assessing spatial and socioeconomic inequalities – A case study in Utrecht, the Netherlands \cite{KNAP2023acomposite} & \cellcolor{gray!15}Transportation (Urban Transport and Accessibility) & \cellcolor{gray!15}$25^\circ$&\cellcolor{gray!15}36 \\
&  Understanding the determinants of x-minute city policies: A review of the North American and Australian cities’ planning documents \cite{LU2023understanding}& Transportation (Urban Transport and Accessibility) & $30^\circ$&32 \\
\midrule
\multirow{3}{*}{\textbf{4}} &  The 15-minute city: Urban planning and design efforts toward creating sustainable neighborhoods \cite{KHAVARIANGARMSIR2023the15}& Transportation (Urban Transport and Accessibility) & $5^\circ$&128 \\
& \cellcolor{gray!15} The 15-minute city: interpreting the model to bring out urban resiliencies \cite{ABDELFATTAH2022the15}& \cellcolor{gray!15}Transportation (Urban Transport and Accessibility) & \cellcolor{gray!15}$10^\circ$&\cellcolor{gray!15}86 \\
&  The 15-minute city for all? – Measuring individual and temporal variations in walking accessibility \cite{WILLBERG2023the15}& Transportation (Urban Transport and Accessibility) & $11^\circ$&76 \\
\midrule
\multirow{3}{*}{\textbf{5}} &  Barcelona under the 15-Minute City Lens: Mapping the Accessibility and Proximity Potential Based on Pedestrian Travel Times \cite{ferrer2022barcelona}& Transportation (Urban Transport and Accessibility) & $8^\circ$&98 \\
& \cellcolor{gray!15} Urban Transition and the Return of Neighbourhood Planning. Questioning the Proximity Syndrome and the 15-Minute City \cite{marchigiani2022urban} & \cellcolor{gray!15}Management, Monitoring, Policy and Law (Urban Planning and Valuation) & \cellcolor{gray!15}$26^\circ$&\cellcolor{gray!15}34 \\
&  Pathways to 15-Minute City adoption: Can our understanding of climate policies' acceptability explain the backlash towards x-minute city programs? \cite{MARQUET2024pathways}& Communication (Social Media and Politics) & $62^\circ$&13 \\
\midrule

\bottomrule
\end{tabular}
}
\caption{Top Root Set Papers by citations by Cluster}\label{tab:citations}
\end{table}

Finally, table~\ref{tab:centrality-measures} shows the top five authors based on five centrality metrics in the co-authorship network: degree, betweenness, closeness, eigenvector, and PageRank. Using multiple metrics helps to capture different aspects of influence and collaboration patterns that would not be visible by looking only at the number of co-authors.
For example, B. Büttner ranks highly across nearly all measures, suggesting not only that he collaborates widely but also that he holds a central and structurally important position in the network. In contrast, authors like K. Geurs or G. Pozoukidou appear only in the betweenness ranking. This indicates that their influence comes from acting as connectors between otherwise separate groups, rather than from collaborating with many others directly.
The eigenvector and PageRank metrics offer yet another perspective: they highlight authors who are connected to other well-connected or central collaborators. Among the top 5 authors by PageRank, Z. Allam and C. Moreno emerge, reflecting their foundational contributions to the development of the topic, which has since attracted significant attention and collaboration from other leading researchers in the field.
Overall, the centrality measures reveal different dimensions of academic presence within a collaborative network allowing a more complete and nuanced understanding than simply counting the number of collaborations.

We can compare the research items collected and organized through \methodname{} to a systematic literature review such as~\cite{papadopoulos2023measuring}. In that study, the authors identified 78 conference and journal research papers from Scopus using the keywords “15 min* cit*” and “20 min* cit*”, without specifying a time range. After filtering for English-language papers and applying full-text eligibility criteria, they retained 40 articles. Subsequently, they conducted both backward and forward snowballing (i.e., identifying additional relevant works by examining the references of the selected articles and the papers that cited them) resulting in a final corpus of 45 articles. In contrast, using \methodname{} with the query “15 minut* city” and restricting the search to articles published from January 1, 2019 onward, we retrieved 409 root set articles, of which 285 were journal or conference papers written in English. Including the base set, the total rises to 2615 articles. Notably, we find that our methodology is able to collect a more conspicuous number of articles; of course, the list of papers presented in the cited work is the result of an extensive process of manual selection, where papers were carefully curated and evaluated to ensure relevance and quality. Indeed, the construction of a literature review does not depend solely on the number of articles retrieved through keyword-based queries; however, a very large volume of results can significantly complicate the process, making a systematic review on a given topic both extensive and time-consuming. To address this challenge, \methodname{} provides mechanisms to help researchers prioritize the most relevant articles. This can be achieved by ranking articles based on centrality measures within the citation network, or by identifying articles authored by individuals who themselves occupy structurally central positions in the co-authorship network. Furthermore, the availability of topic modeling enables filtering of the corpus based on thematic content, while the application of community detection algorithms allows for the identification of coherent groups of articles or authors. 
These communities can then be used to focus the analysis on specific subfields or research clusters, thereby supporting a more targeted and in-depth exploration of the scientific landscape. Furthermore, the articles identified through \methodname{}—which can be exported in CSV format—are readily filterable by language, document type (e.g., journal and conference papers only), and inclusion or exclusion of the base set. This significantly streamlines the work of researchers who need to identify, organize, and evaluate relevant articles for conducting a systematic review.

\begin{table}[h]
\centering
\footnotesize
\begin{tabular}{@{}llcllcllc@{}}
\toprule
\textbf{Rank} & \multicolumn{2}{l}{\textbf{Degree}} & \multicolumn{2}{l}{\textbf{Betweenness}} & \multicolumn{2}{l}{\textbf{Closeness}} \\ \midrule
1 & B. Büttner        & \texttt{1.40e-1} & B. Büttner       & \texttt{1.10e-2} & B. Büttner         & \texttt{1.40e-1} \\
2 & M. T. B. Larriva  & \texttt{1.03e-1} & K. Geurs         & \texttt{1.87e-3} & A. Gorrini         & \texttt{1.12e-1} \\
3 & J. Nofre          & \texttt{1.02e-1} & G. Pozoukidou    & \texttt{9.51e-4} & D. Presicce        & \texttt{1.12e-1} \\
4 & J. Carpio-Pinedo  & \texttt{1.02e-1} & J. Nofre         & \texttt{6.34e-4} & F. Messa           & \texttt{1.12e-1} \\
5 & A. Gorrini        & \texttt{1.02e-1} & L. Abdelfattah   & \texttt{6.34e-4} & M. T. B. Larriva   & \texttt{1.12e-1} \\
\midrule
\textbf{Rank} & \multicolumn{2}{l}{\textbf{Eigenvector}} & \multicolumn{2}{l}{\textbf{PageRank}} & \multicolumn{2}{l}{~} \\ \midrule
1 & B. Büttner        & \texttt{1.01e-1} & Z. Allam         & \texttt{3.24e-3} & ~ & ~ \\
2 & M. T. B. Larriva  & \texttt{1.00e-1} & B. Büttner       & \texttt{2.98e-3} & ~ & ~ \\
3 & A. Gorrini        & \texttt{1.00e-1} & C. Moreno        & \texttt{2.51e-3} & ~ & ~ \\
4 & D. Presicce       & \texttt{1.00e-1} & B. Murgante      & \texttt{2.49e-3} & ~ & ~ \\
5 & F. Messa          & \texttt{1.00e-1} & O. Marquet       & \texttt{2.40e-3} & ~ & ~ \\
\bottomrule
\end{tabular}
\caption{Top 5 authors by various centrality measures}
\label{tab:centrality-measures}
\end{table}

\section{Conclusion, Limitations and Future Work}

\methodname{} is a Python package designed to easily collect bibliographic data on a specific query and to perform bibliometric analyses based on network science. To do so, it includes a number of methods that perform all the required steps to parse the query, collect and export the bibliographic data, build and analyze the citation and co-authorship graphs. While it currently relies on OpenAlex to gather bibliometric data, the modular structure of \methodname{} allows users to include their own set of functions to include further sources of data, such as the Elsevier Research Products APIs\footnote{https://dev.elsevier.com/} or PubMed APIs\footnote{https://www.ncbi.nlm.nih.gov/home/develop/api/}, as long as the final outputs of this module are conveniently structured for the ingestion by the module that performs the bibliometric analysis. 

\methodname{} is particularly helpful in exploring a strongly multidisciplinary scientific domain, as seen in the use case of Sec.~\ref{sec:usecase}. Basing the bibliometric analysis on network analysis measures is effective since it allows us to assess the structural relationships between publications, authors, and research topics. By leveraging network analysis measures, \methodname{} enables the identification of key authors, influential publications, and emerging research clusters. This approach provides insights into collaborative patterns, disciplinary intersections, and the evolution of knowledge within the scientific domain, revealing trends and influential nodes that traditional bibliometric methods may overlook. Moreover, in such a multidisciplinary environment, using this technique helps a researcher navigate an unfamiliar domain.

Notably, we find that our methodology is able to collect a more conspicuous number of diverse sources; of course, the list of papers presented in a systematic literature review is the result of an extensive process of manual selection, where papers were carefully curated and evaluated to ensure relevance and quality. In contrast, \methodname{} serves as a complementary tool positioned at the initial stages of a researcher’s workflow, designed to bring a broad range of works into consideration with minimal effort. By leveraging network analysis, \methodname{} is capable of identifying influential publications across a diverse array of topics that are relevant to the query. This approach minimizes biases that could stem from the researcher's preconceptions, promoting a more comprehensive and impartial exploration of the field. Consequently, it enhances the researcher’s ability to access a wider and potentially richer pool of resources, which can then be subjected to finer, manual selection processes later on. This combination of automated and manual review processes optimizes the rigor and breadth of a literature review, fostering a more heterogeneous view of the scientific landscape.

Nevertheless, it is important to acknowledge some limitations of the current implementation of \methodname{}. Specifically, the analysis is restricted to citation and co-authorship networks. Other network structures could be explored, such as author-to-author citation networks. Moreover, we employed a selected set of centrality measures and community detection algorithms, but incorporating a broader range of metrics and community detection algorithms could enhance the depth of the bibliometric analysis. Additionally, the interoperability between the collection module and other tools or libraries in Python and R could be improved by supporting export in standardized formats, such as those used by Web of Science, Dimensions,
or PubMed, which are commonly accepted as input by many bibliometric analysis tools. Furthermore, the topic modeling module is currently quite basic, and it does not implement any caching mechanisms, which could significantly affect performance on large datasets. Finally, the generated visualizations are static and offer limited customization; introducing interactive and configurable visual outputs would make the tool more flexible and accessible to a broader range of users. We plan to fix some of these limitations in future versions of the library.

\begin{appendices}

\section{Documentation of the \texttt{openalex} Module}
\label{appendixA}

This appendix provides documentation for the \texttt{openalex} module, detailing its public methods and their functionalities.

\subsection{Core Functions}

\subsubsection{\texttt{string\_generator\_from\_lite\_regex}}

Generates a list of query strings from a simplified regular expression pattern.

\textbf{Parameters:}
\begin{itemize}
    \item \texttt{queries\_regex} (str): A string representing a simplified regex pattern.
\end{itemize}

\textbf{Returns:}
\begin{itemize}
    \item \texttt{List[str]}: List of query strings.
\end{itemize}

\subsubsection{\texttt{query\_OpenAlex}}

Queries the OpenAlex API based on the given parameters.

\textbf{Parameters:}
\begin{itemize}
    \item \texttt{api\_type} (str): Type of API query ('search', 'cite', 'cited\_by').
    \item \texttt{parameter} (Optional[str]): Article ID for 'cite' and 'cited\_by' queries.
    \item \texttt{mail} (str): Email for API request identification.
    \item \texttt{from\_publication\_date} (Optional[str]): Start date for filtering (YYYY-MM-DD).
    \item \texttt{to\_publication\_date} (Optional[str]): End date for filtering (YYYY-MM-DD).
    \item \texttt{cache} (bool): Use caching.
\end{itemize}

\textbf{Returns:}
\begin{itemize}
    \item \texttt{List[Dict[str, Dict]]}: List of results matching the query.
\end{itemize}

\subsubsection{\texttt{retrieve\_articles}}
Retrieves root set articles from OpenAlex based on a list of queries and optional parameters.

\textbf{Parameters:}
\begin{itemize}
    \item \texttt{queries} (List[str]): Search queries.
    \item \texttt{mail} (str): Email for API request identification.
    \item \texttt{from\_publication\_date} (Optional[str]): Start date (YYYY-MM-DD).
    \item \texttt{to\_publication\_date} (Optional[str]): End date (YYYY-MM-DD).
\end{itemize}

\textbf{Returns:}
\begin{itemize}
    \item \texttt{str}: File path to the JSON results. 
\end{itemize}

\subsubsection{\texttt{export\_articles\_to\_csv}}
Exports articles data from a JSON file to CSV.

\textbf{Parameters:}
\begin{itemize}
    \item \texttt{json\_file\_path} (str): Path to the JSON file.
    \item \texttt{fields\_to\_export} (Optional[List[str]]): Fields to include.
    \item \texttt{export\_path} (Optional[str]): Path for the exported CSV.
\end{itemize}

\textbf{Returns:}
\begin{itemize}
    \item \texttt{None}: Save selected fields to a CSV file.  
\end{itemize}

\subsubsection{\texttt{export\_authors\_to\_csv}}
Exports author data from a JSON file to CSV.

\textbf{Parameters:}
\begin{itemize}
    \item \texttt{json\_file\_path} (str): Path to the JSON file.
    \item \texttt{fields\_to\_export} (Optional[List[str]]): Fields to include.
    \item \texttt{export\_path} (Optional[str]): Path for the exported CSV.
\end{itemize}

\textbf{Returns:}
\begin{itemize}
    \item \texttt{None}: Save selected fields to a CSV file.  
\end{itemize}

\subsubsection{\texttt{export\_institutions\_to\_csv}}
Exports institution data from a JSON file to CSV.

\textbf{Parameters:}
\begin{itemize}
    \item \texttt{json\_file\_path} (str): Path to the JSON file.
    \item \texttt{fields\_to\_export} (Optional[List[str]]): Fields to include.
    \item \texttt{export\_path} (Optional[str]): Path for the exported CSV.
\end{itemize}
\textbf{Returns:}
\begin{itemize}
    \item \texttt{None}: Save selected fields to a CSV file.  
\end{itemize}

\subsubsection{\texttt{export\_venues\_to\_csv}}
Exports venue data from a JSON file to CSV.

\textbf{Parameters:}
\begin{itemize}
    \item \texttt{json\_file\_path} (str): Path to the JSON file.
    \item \texttt{fields\_to\_export} (Optional[List[str]]): Fields to include.
    \item \texttt{export\_path} (Optional[str]): Path for the exported CSV.
\end{itemize}
\textbf{Returns:}
\begin{itemize}
    \item \texttt{None}: Save selected fields to a CSV file.  
\end{itemize}

\subsubsection{\texttt{export\_articles\_to\_scopus}}
Exports author data from a JSON file to Scopus like CSV.

\textbf{Parameters:}
\begin{itemize}
    \item \texttt{json\_file\_path} (str): Path to the JSON file.
    \item \texttt{include\_periphery} (Optional[boolean]): Include the base set.
    \item \texttt{export\_path} (Optional[str]): Path for the exported CSV.
\end{itemize}
\textbf{Returns:}
\begin{itemize}
    \item \texttt{None}: Save selected fields to a CSV file.  
\end{itemize}

\subsubsection{\texttt{create\_citation\_graph}}
Creates a directed graph from article data representing citations.

\textbf{Parameters:}
\begin{itemize}
    \item \texttt{articles} (Dict[str, Dict]): Article data.
    \item \texttt{export\_path} (Optional[str]): Path to export the graph.
\end{itemize}

\textbf{Returns:}
\begin{itemize}
    \item \texttt{nx.DiGraph}: An undirected co-authorship graph.  
\end{itemize}
\subsubsection{\texttt{create\_coauthorship\_graph}}
Creates an undirected graph representing co-authorship relationships.

\textbf{Parameters:}
\begin{itemize}
    \item \texttt{articles} (Dict[str, Dict]): Article data.
    \item \texttt{export\_path} (Optional[str]): Path to export the graph.
    \item \texttt{periphery} (bool): Include base set articles.
\end{itemize}

\textbf{Returns:}
\begin{itemize}
    \item \texttt{nx.Graph}: An undirected co-authorship graph.  
\end{itemize}

\section{Documentation of the \texttt{bibliometric\_analysis} Module}
\label{appendixB}
\subsection{Core Functions}
\subsubsection{\texttt{extract\_metrics\_to\_csv}}
Extracts the specified metrics and fields from the graph and saves them to a CSV file.

\textbf{Parameters:}
\begin{itemize}
    \item \texttt{G} (\texttt{nx.DiGraph|nx.Graph}): A NetworkX directed or undirected graph.
    \item \texttt{metrics} (\texttt{List[str]}): A list of metrics to compute (available: 'betweenness\_centrality', 'closeness\_centrality', 'page\_rank', 'in\_degree', 'out\_degree', 'degree').
    \item \texttt{fields} (\texttt{List[str]}): A list of fields to include in the CSV.
    \item \texttt{csv\_file\_path} (\texttt{str}): The path to the CSV file to save the results.
\end{itemize}

\textbf{Returns:}
\begin{itemize}
    \item \texttt{None}: Save selected fields and clusters to a CSV file.  
\end{itemize}

\subsubsection{\texttt{extract\_clusters\_to\_csv}}
Extracts the cluster and selected fields from the graph and saves them to a CSV file.

\textbf{Parameters:}
\begin{itemize}
    \item \texttt{G} (\texttt{nx.DiGraph|nx.Graph}): A NetworkX directed or undirected graph.
    \item \texttt{fields} (\texttt{List[str]}): A list of fields to include in the CSV.
    \item \texttt{csv\_file\_path} (\texttt{str}): The path to the CSV file to save the results.
\end{itemize}
\textbf{Returns:}
\begin{itemize}
    \item \texttt{None}: Save selected fields and metrics to a CSV file.  
\end{itemize}

\subsubsection{\texttt{cluster\_graph}}
Clusters the nodes in the graph using the specified clustering algorithm and returns a new graph with cluster numbers as node attributes.

\textbf{Parameters:}
\begin{itemize}
    \item \texttt{G} (\texttt{nx.DiGraph|nx.Graph}): A NetworkX directed or undirected graph.
    \item \texttt{algorithm} (\texttt{str}): The clustering algorithm to use
    (\texttt{int}): The number of clusters to create when Spectral Clustering is used.
\end{itemize}

\textbf{Returns:}
\begin{itemize}
    \item A NetworkX graph with cluster numbers as a new node attribute.
\end{itemize}

\subsection{Charts Functions}

\subsubsection{\texttt{plot\_article\_trends}}
Generates a stacked area chart visualizing the number of root set and base set articles over time, aggregated by the specified interval (\texttt{month}, \texttt{quarter}, \texttt{year}).

\textbf{Parameters:}
\begin{itemize}
    \item \texttt{articles} (\texttt{dict}): A dictionary containing articles with metadata.
    \item \texttt{interval} (\texttt{str}): Time interval for the x-axis labels. Can be \texttt{"month"}, \texttt{"quarter"}, or \texttt{"year"}.
    \item \texttt{color\_core} (\texttt{Optional[str]}): Color for the root set articles in the plot.
    \item \texttt{color\_periphery} (\texttt{Optional[str]}): Color for the base set articles.
    \item \texttt{date\_from} (\texttt{Optional[datetime]}): Start date for the aggregation range.
    \item \texttt{date\_to} (\texttt{Optional[datetime]}): End date for the aggregation range.
    \item \texttt{num\_ticks} (\texttt{Optional[int]}): Maximum number of x-axis ticks to display.
\end{itemize}

\textbf{Returns:}
\begin{itemize}
    \item \texttt{None}: The function generates and displays the stacked area chart.
\end{itemize}

\subsubsection{\texttt{plot\_topic\_trends}}
Generates a grouped stacked bar chart to visualize aggregated data for topics or concepts over time. Articles can be split into root set and base set articles, with the option to filter by specific topics or concepts.

\textbf{Parameters:}
\begin{itemize}
    \item \texttt{articles} (\texttt{dict}): A dictionary containing articles with metadata.
    \item \texttt{interval} (\texttt{Optional[str]}): Aggregation interval for the x-axis (\texttt{"month"}, \texttt{"quarter"}, \texttt{"year"}).
    \item \texttt{top\_n\_colors} (\texttt{Optional[list]}): List of colors to use for the bars.
    \item \texttt{show\_periphery} (\texttt{Optional[bool]}): If \texttt{True}, display base set articles.
    \item \texttt{show\_core} (\texttt{Optional[bool]}): If \texttt{True}, display root set articles.
    \item \texttt{field\_key} (\texttt{Optional[str]}): Key to filter by for topics (\texttt{"field"} or \texttt{"domain"}).
    \item \texttt{date\_from} (\texttt{Optional[datetime]}): Start date for filtering.
    \item \texttt{date\_to} (\texttt{Optional[datetime]}): End date for filtering.
    \item \texttt{top\_n} (\texttt{Optional[int]}): Number of top topics/concepts to display.
    \item \texttt{num\_ticks} (\texttt{Optional[int]}): Maximum number of ticks on the x-axis.
\end{itemize}

\textbf{Returns:}
\begin{itemize}
    \item \texttt{None}: The function generates and displays the grouped stacked bar chart.
\end{itemize}

\subsubsection{\texttt{plot\_top\_authors}}
Generates a stacked bar chart visualizing the top authors and their articles, split by topics or concepts. It can rank authors by the number of citations or publications.

\textbf{Parameters:}
\begin{itemize}
    \item \texttt{articles} (\texttt{dict}): A dictionary containing articles with metadata.
    \item \texttt{field\_key} (\texttt{Optional[str]}): Key to filter topics by (\texttt{"field"} or \texttt{"domain"}).
    \item \texttt{date\_from} (\texttt{Optional[datetime]}): Start date for filtering.
    \item \texttt{date\_to} (\texttt{Optional[datetime]}): End date for filtering.
    \item \texttt{num\_authors} (\texttt{int}): Number of top authors to display.
    \item \texttt{by\_citations} (\texttt{bool}): If \texttt{True}, ranks authors by the number of citations instead of publications.
    \item \texttt{show\_periphery} (\texttt{bool}): If \texttt{True}, includes base set articles in the analysis.
    \item \texttt{show\_core} (\texttt{bool}): If \texttt{True}, includes root set articles in the analysis.
    \item \texttt{n\_colors} (\texttt{Optional[list]}): List of colors for the chart bars.
\end{itemize}

\textbf{Returns:}
\begin{itemize}
    \item \texttt{None}: The function generates and displays the stacked bar chart.
\end{itemize}

\subsubsection{\texttt{plot\_top\_keywords\_from\_abstracts}}
Generates a chart to visualize the most frequent keywords extracted from article abstracts.

\textbf{Parameters:}
\begin{itemize}
    \item \texttt{articles} (\texttt{dict}): A dictionary containing articles with metadata.
    \item \texttt{date\_from} (\texttt{Optional[datetime]}): Start date for filtering.
    \item \texttt{date\_to} (\texttt{Optionaldatetime]}): End date for filtering.
    \item \texttt{show\_core} (\texttt{bool}): If \texttt{True}, includes root set articles in the analysis.
    \item \texttt{show\_periphery} (\texttt{bool}): If \texttt{True}, includes base set articles in the analysis.
    \item \texttt{top\_n} (\texttt{int}): Number of top keywords to display.
    \item \texttt{ngram\_range} (\texttt{tuple}): N-gram range for extracting keywords.
\end{itemize}

\textbf{Returns:}
\begin{itemize}
    \item \texttt{None}: The function generates and displays the chart.
\end{itemize}

\subsubsection{\texttt{plot\_keyword\_trends}}
Generates a chart to visualize the trends of top keywords over time.

\textbf{Parameters:}
\begin{itemize}
    \item \texttt{articles} (\texttt{dict}): A dictionary containing articles with metadata.
    \item \texttt{date\_from} (\texttt{Optional[datetime]}): Start date for filtering.
    \item \texttt{date\_to} (\texttt{Optional[datetime]}): End date for filtering.
    \item \texttt{show\_core} (\texttt{bool}): If \texttt{True}, includes root set articles in the analysis.
    \item \texttt{show\_periphery} (\texttt{bool}): If \texttt{True}, includes base set articles in the analysis.
    \item \texttt{top\_n} (\texttt{int}): Number of top keywords to display.
    \item \texttt{ngram\_range} (\texttt{tuple}): N-gram range for extracting keywords.
    \item \texttt{interval} (\texttt{str}): Time interval for aggregation (\texttt{"month"}, \texttt{"quarter"}, \texttt{"year"}).
     \item \texttt{n\_cluster}(\texttt{Optional[List[str]]}): List of colors
\end{itemize}

\textbf{Returns:}
\begin{itemize}
    \item \texttt{None}: The function generates and displays the chart.
\end{itemize}

\subsection{Charts Networks Functions}
\subsubsection{\texttt{show\_clustered\_graph}}
Visualizes clusters in a graph with pie charts representing the distribution of entries within each cluster. It also allows exporting article information in a CSV file.

\textbf{Parameters:}
\begin{itemize}
    \item \texttt{G} (\texttt{nx.DiGraph}): A NetworkX directed graph containing nodes with cluster data.
    \item \texttt{n\_clusters} (\texttt{Optional[int]}): The number of top clusters to display.
    \item \texttt{m\_entries} (\texttt{Optional[int]}): The number of top entries to display in pie charts.
    \item \texttt{n\_cluster\_colors} (\texttt{Optional[List[str]]}): List of colors for clusters. 
    \item \texttt{m\_entry\_colors} (\texttt{Optional[List[str]]}): List of colors for pie chart entries. 
    \item \texttt{min\_node\_radius} (\texttt{Optional[float]}): Minimum radius for the nodes.
    \item \texttt{max\_node\_radius} (Optional[float]): Maximum radius for the nodes.
    \item \texttt{min\_pie\_radius} (Optional[float]): Minimum radius for pie charts.
    \item \texttt{max\_pie\_radius} (Optional[float]): Maximum radius for pie charts.
    \item \texttt{size\_legend\_marker} (Optional[int]): Marker size for the legend.
    \item \texttt{size\_legend\_font} (Optional[int]): Font size for the legend.
    \item \texttt{size\_node\_font} (Optional[int]): Font size for node annotations.
    \item \texttt{min\_edge\_width} (Optional[int]): Minimum width of edges between clusters.
    \item \texttt{max\_edge\_width} (Optional[int]): Maximum width of edges between clusters.
    \item \texttt{edge\_color} (Optional[str]): Color for the edges.
    \item \texttt{top\_m\_entries} (\texttt{Optional[List[str]]}): List of top entries to display.
    \item \texttt{bbox\_to\_anchor\_legend\_entries} (\texttt{Optional[Tuple[float, float, float, float]]}): Bounding box for entries legend.
    \item \texttt{bbox\_to\_anchor\_legend\_clusters} (\texttt{Optional[Tuple[float, float, float, float]]}): Bounding box for clusters legend. 
    \item \texttt{image\_size} (\texttt{Optional[Tuple[int, int]]}): Size of the output image.
    \item \texttt{topics\_level} (\texttt{Optional[str]}): Level of topics to consider (\texttt{"field"} or \texttt{"domain"}).
    \item \texttt{export\_path\_png} (\texttt{Optional[str]}): Path to export the figure as PNG.
\end{itemize}

\textbf{Returns:}
\begin{itemize}
    \item \texttt{None}: The function generates and displays the clustered graph with pie charts.
\end{itemize}

\subsubsection{\texttt{show\_cluster\_statistics}}
Displays a horizontal bar chart showing the number of nodes in each cluster, along with a table containing cluster statistics.

\textbf{Parameters:}
\begin{itemize}
    \item \texttt{csv\_file\_path} (\texttt{str}): Path to the CSV file containing cluster data.
    \item \texttt{color} (\texttt{Optional[str]}): Color for the bar chart.
    \item \texttt{image\_size} (\texttt{Optional[Tuple[int, int]]}): Size of the image for the output plot.
    \item \texttt{n\_clusters} (\texttt{Optional[int]}): Number of top clusters to display.
\end{itemize}

\textbf{Returns:}
\begin{itemize}
    \item \texttt{None}: The function generates and displays the bar chart and statistics table.
\end{itemize}

\subsubsection{\texttt{show\_graph\_statistics}}
Displays a variety of statistics about the graph, including the number of nodes, number of edges, graph density, and various centrality measures. It also generates a correlation matrix and distribution histograms for centrality measures.

\textbf{Parameters:}
\begin{itemize}
    \item \texttt{G} (\texttt{nx.Graph}): The graph whose statistics are to be displayed.
    \item \texttt{csv\_file\_path} (\texttt{str}): Path to the CSV file containing centrality measure data.
    \item \texttt{header\_color} (\texttt{str}, optional): Color for the table headers.
\end{itemize}

\textbf{Returns:}
\begin{itemize}
    \item \texttt{None}: The function generates and displays various statistics, including a table, correlation heatmap, and histograms.
\end{itemize}

\end{appendices}


\bibliography{bibliography.bib}

\end{document}